\documentclass[fleqn,usenatbib]{mnras}

\usepackage{newtxtext,newtxmath}
\usepackage[T1]{fontenc}
\usepackage{ae,aecompl}

\usepackage{graphicx}	% Including figure files
\usepackage{amsmath}	% Advanced maths commands
\usepackage{amssymb}	% Extra maths symbols

\usepackage{verbatim}
\usepackage{multirow}
\usepackage{threeparttable}
\usepackage{relsize}
\usepackage{color}
\def\ga{\mathrel{\hbox{\rlap{\hbox{\lower4pt\hbox{$\sim$}}}\hbox{$>$}}}}
\def\la{\mathrel{\hbox{\rlap{\hbox{\lower4pt\hbox{$\sim$}}}\hbox{$<$}}}}

\usepackage[normalem]{ulem}

\definecolor{darkgreen}{rgb}{0.13, 0.55, 0.13}
\definecolor{orange}{rgb}{0.8, 0.15, 0.13}

\newcommand{\aref}[1]{\hyperref[#1]{Appendix~\ref{#1}}}

\usepackage{scalerel,tikz}
\usetikzlibrary{svg.path}
\definecolor{orcidlogocol}{HTML}{A6CE39}
\tikzset{orcidlogo/.pic={
 \fill[orcidlogocol] svg{M256,128c0,70.7-57.3,128-128,128C57.3,256,0,198.7,0,128C0,57.3,57.3,0,128,0C198.7,0,256,57.3,256,128z};
 \fill[white] svg{M86.3,186.2H70.9V79.1h15.4v48.4V186.2z}
 svg{M108.9,79.1h41.6c39.6,0,57,28.3,57,53.6c0,27.5-21.5,53.6-56.8,53.6h-41.8V79.1z M124.3,172.4h24.5c34.9,0,42.9-26.5,42.9-39.7c0-21.5-13.7-39.7-43.7-39.7h-23.7V172.4z}
 svg{M88.7,56.8c0,5.5-4.5,10.1-10.1,10.1c-5.6,0-10.1-4.6-10.1-10.1c0-5.6,4.5-10.1,10.1-10.1C84.2,46.7,88.7,51.3,88.7,56.8z};
}}
\newcommand\orcidicon[1]{\href{https://orcid.org/#1}{\mbox{\scalerel*{
\begin{tikzpicture}[yscale=-1,transform shape]
\pic{orcidlogo};
\end{tikzpicture}
}{|}}}}

\title[Biases in measurements of cloud kinematics]{Understanding biases in measurements of molecular cloud kinematics using line emission}

\author[Yuan et al.]{Yuxuan Yuan,$^{1}$~\orcidicon{0000-0001-6816-0682}\thanks{E-mail: yuxuan@mso.anu.edu.au (YY)}
Mark R. Krumholz$^{1,2}$~\orcidicon{0000-0003-3893-854X},
Blakesley Burkhart$^{3,4}$
\\
$^{1}$Research School of Astronomy and Astrophysics, Australian National University, Canberra, ACT 2601, Australia\\
$^{2}$ARC Centre of Excellence for Astronomy in Three Dimensions (ASTRO-3D), Canberra, ACT 2601, Australia\\
$^{3}$Center for Computational Astrophysics, Flatiron Institute, 162 Fifth Avenue, New York, NY 10010, USA\\
$^{4}$Department of Physics and Astronomy, Rutgers University, 136 Frelinghuysen Rd, Piscataway, NJ 08854, USA
}

\date{Accepted XXX. Received YYY; in original form ZZZ}

\pubyear{2020}

\begin{document}
\label{firstpage}
\pagerange{\pageref{firstpage}--\pageref{lastpage}}
\maketitle

\begin{abstract}
Molecular line observations using a variety of tracers are often used to investigate the kinematic structure of molecular clouds. However, measurements of cloud velocity dispersions with different lines, even in the same region, often yield inconsistent results. The reasons for this disagreement are not entirely clear, since molecular line observations are subject to a number of biases. In this paper, we untangle and investigate various factors that drive linewidth measurement biases by constructing synthetic position-position-velocity cubes for a variety of tracers from a suite of self-gravitating magnetohydrodynamic simulations of molecular clouds. We compare linewidths derived from synthetic observations of these data cubes to the true values in the simulations. We find that differences in linewidth as measured by different tracers are driven by a combination of density-dependent excitation, whereby tracers that are sensitive to higher densities sample smaller regions with smaller velocity dispersions, opacity broadening, especially for highly optically thick tracers such as CO, and finite resolution and sensitivity, which suppress the wings of emission lines. 
We find that, at fixed signal to noise ratio, three commonly-used tracers, the $J=4\rightarrow 3$ line of CO, the $J=1\rightarrow 0$ line of C$^{18}$O and the $(1,1)$ inversion transition of NH$_3$, generally offer the best compromise between these competing biases, and produce estimates of the velocity dispersion that reflect the true kinematics of a molecular cloud to an accuracy of $\approx 10\%$ regardless of the cloud magnetic field strengths, evolutionary state, or orientations of the line of sight relative to the magnetic field. Tracers excited primarily in gas denser than that traced by NH$_3$ tend to underestimate the true velocity dispersion by $\approx 20\%$ on average, while low-density tracers that are highly optically thick tend to have biases of comparable size in the opposite direction.
\end{abstract}
%%%%%%%%%%%%%%%%%%%%%%%%%%%%%%%%%%%%%%%%%%%%%%%%%%

%%%%%%%%%%%%%%%%% BODY OF PAPER %%%%%%%%%%%%%%%%%%

\section{Introduction}
\label{intro}

Molecular line emission is one of our primary tools for characterizing the dense interstellar medium. Line observations are uniquely rich in that they carry information not just on the location of gas, but on its physical properties and kinematics. In particular, the velocity information provided by lines allows one to compute the mean velocity, the velocity dispersion, and a variety of higher-order statistics along each line of sight. The brightest molecular lines in the Milky Way and nearby galaxies are the first few rotational lines of CO, and it has long been known that the dispersion of the CO line is much larger than would be expected due to thermal broadening alone, indicating the presence of supersonic motions \citep[e.g.,][]{Liszt74a, Goldreich74a}. Subsequent exploration showed that the linewidth increases systematically with the size of the region probed \citep[e.g.,][]{Larson81a, Solomon87a, Goodman98b,  Bolatto08a}, and that the difference in velocity (measured either as the difference in first velocity moments, or via the $L_2$ norm or a similar norm for the difference in the full spectra) between lines of sight increases systematically with the separation of the sightlines on the plane of the sky \citep[e.g.,][]{Issa90a, Ossenkopf02a, Burkhart09a}. Collectively these correlations are known as the linewidth-size relation.

While the statistics of the CO line have been explored most extensively, similar large velocity spreads are also observed in many other molecular lines, including isotopologues of CO, and a variety of tracers that, for reasons of either chemistry or excitation, are more sensitive to gas denser than that traced by CO. Examples of the latter include the rotational lines of molecules such as HCN, CS, and N$_2$H$^+$, and inversion transitions of molecules such as NH$_3$. These molecules often show different linewidths, and different linewidth-size relations, from CO, even when both are observed along the same line of sight \citep[e.g.,][]{Onishi96a, Goodman98b, Walsh04a, Andre07a, Kirk07a, Muench07a, Rosolowsky08a}.

There have been only limited theoretical attempts to understand the relationships between the kinematics revealed by different tracers. In some cases authors have modelled the kinematics of particular systems observed in multiple tracers \citep[e.g.,][]{Walker-Smith13a, Maureira17a}, but there have been few more general explorations. Consequently, it is not entirely certain what drives the differences between tracers. For example, \citet{Hacar16a} argue that CO linewidths are larger than those seen in rarer isotopologues because opacity broadening artificially inflates the linewidth, causing flows that are actually transsonic to appear supersonic in the CO lines.
However, earlier studies showed that opacity broadening of CO is not a major correction factor for measurements of the sonic Mach number \citep{Correia14a} from linewidths but can be very important for measurements of the Mach number from the density spatial power spectrum \citep{Burkhart13b}.
\citet{Offner08a} argue that density-dependent excitation effects explain the differences in kinematics measured with mostly optically thin tracers such as NH$_3$, N$_2$H$^+$, and C$^{18}$O. The problem is fundamentally difficult because the observed line emission is a complex product of many factors including the underlying gas distribution and kinematics, subtle excitation and radiative transfer effects, and  finite resolution, sensitivity and beam-smearing from the telescopes. All of these effects are difficult to study because they are entangled.

Our goal in this paper is to untangle the factors that drive differences in the kinematics as measured with a range of tracers. Our approach is to rely on simulations and simulated line emission. The great advantage of using simulations is that we precisely know the true underlying kinematics, and we can conduct numerical experiments that would not be possible in reality, for example separating the effects of excitation and opacity by independently turning them on and off. To this end,
in this paper we use a series of simulations of star-formation in a self-gravitating, magnetised, turbulent medium to model line observations for five tracers: CO,C$^{18}$O, HCN, NH$_3$, and N$_2$H$^+$.
We create synthetic position-position-velocity cubes for each, and then analyse the statistical properties of the resulting data. We use these synthetic data to untangle what drives tracer-dependent kinematics.

The structure of this paper is as follows. In \autoref{sec: methods} we describe the numerical simulations and methods we use. We present our results in \autoref{sec: results}, where we find that higher density tracers trace smaller regions and lower linewidth due to the linewidth-size relation. We discuss general findings on which tracers perform best in \autoref{sec: disc}, and give our conclusions in \autoref{con}.

\section{Methods}
\label{sec: methods}

We perform our analysis on a suite of Enzo simulations that we describe in \autoref{ssec:simulations} \citep{Collins12a}. These simulations are part of the Catalog for Astrophysical Turbulence Simulations (CATS) and are publicly available (Burkhart et al. 2020, in prep).  In order to produce synthetic PPV cubes from these simulations, we generate a table of large velocity gradient (LVG) models with the code Derive the Energetics and Spectra of Optically Thick Interstellar Clouds (\textsc{despotic}; \citealt{Krumholz14b}). We describe our method for producing these tables, and for using them to generate PPV cubes, in \autoref{ssec:despotic}. We then describe how we model the effects of finite telescope resolution and signal to noise ratio on these PPV cubes in \autoref{ssec: model_telescope}. The source code and data used in this paper are available from \url{https://github.com/yyx319/Biases-in-measurements-of-cloud-kinematics}

\subsection{Simulations}
\label{ssec:simulations}
We use a suite of three simulations of self-gravitating, isothermal, magnetised gas in a periodic domain performed with the adaptive mesh refinement (AMR) code \textsc{enzo} (see \citealt{Bryan14a} for a general description of the code, and \citealt{Collins12a} for a description of the MHD method). The initial conditions for all three were generated by a suite of unigrid simulations using the \textsc{ppml} code \citep{Ustyugov09}  without self-gravity.
These simulation are described in detail in \citet{Collins12a}.  The initial conditions include a uniform density field and magnetic field initialized along a preferred direction. The box is driven with a pure solenoidal pattern until a steady turbulent state is reached.
At the end of the stirring phase, all three simulations have fully-developed turbulence with virial parameter
\begin{equation}
\alpha_{\rm vir} = \frac{5 v_{\rm rms}^2}{3 G \rho_0 L_0^2} = 1    
\end{equation}
and sonic Mach number $\mathcal{M}_s=v_{\rm rms}/c_s = 9$, where $\rho_0$ is the mean density in the simulation box, $L_0$ is the size of the box, $c_s$ is the isothermal sound speed, and $v_{\rm rms}$ is the root mean square velocity. The three simulations have plasma $\beta$ values $\beta_0 = 0.2$, 2.0, and 20.0, respectively.

Once statistical steady state is reached, gravity is turned on and the simulations are allowed to evolve with no further driving. We study snapshots from $t=0$ to $t=0.6t_{\rm ff}$ after gravity is turned on. During the self-gravitating phase, the root grid resolution is $512^3$, and we add on top of this four levels of refinement by a factor of two. The refinement condition is such that the local Jeans length $L_{\rm J} =\sqrt{c_s^2\pi/G\rho}$ is always resolved by at least 16 zones. This gives an effective linear resolution of 8192.

Isothermal self-gravitating flows of the type used in our simulation suite can be re-scaled to vary the gas density, length, and other parameters (see \autoref{ssec:density_dependence} for further discussion), but in order to calculate the observable emission we need to choose a particular set of physical values of the various simulation parameters. We therefore adopt the following fiducial scalings, which are typical of observed molecular clouds in the Milky Way:

\begin{eqnarray} 
t_{\rm ff}& = &1.1\;\mbox{Myr}\\
L_0& =& 4.6\;\mbox{pc} \\
v_{\rm rms}& = &1.8\;\mbox{km s}^{-1},\\
M &= &5900 \;M_\odot\\
B_0& =& (13, 4.4, 1.3)\;\mu\mbox{G}.
\end{eqnarray}

These choices correspond to adopting $c_s = 0.2$ km s$^{-1}$ and a hydrogen number density $n_{\rm H} = 1000$ cm$^{-3}$. We return to the issue of scaling in \autoref{ssec:density_dependence}.

\subsection{Line emission calculation}
\label{ssec:despotic}
We calculate the observable molecular line luminosity from the simulations using the code \textsc{despotic} \citep{Krumholz14b}. We perform these calculations for the following lines: HCN J $= 1\to 0$, CO J $= 1\to 0$ and J $= 4\to 3$, ${\rm C}^{18}{\rm O}$ J $= 1\to 0$ and J $= 4\to 3$, ${\rm N}_2{\rm H}^+$ J $= 1\to 0$ and NH$_3$ $(1,1)$, as they span a wide range of densities at which they are effectively excited. We are particularly interested in different lines and transitions of CO and its isotopolgues, since these lines are bright and they are often used for wide-field mapping; we use the J $=4\to3$ line as an example that should be representative of transitions at intermediate J in general. We do not include $^{13}$CO as a separate case, because testing shows that the results for it are just intermediate between those for CO and C$^{18}$O.

\textsc{Despotic} solves the equations of statistical equilibrium for the level populations of each species, including non-local thermodynamic equilibrium effects. It uses an escape probability formalism to treat optical depth effects. \textsc{Despotic} implements multiple choices for how to calculate the escape probability, and for this work we use the large velocity gradient (LVG) approximation \citep{Goldreich74a, de-Jong80a}. The details of the numerical method are provided in \citet{Krumholz14b}. We use collision rate and Einstein coefficients taken from the Leiden Atomic and Molecular Database \citep{Schoier05a} for all calculations. The underlying collision rate data for HCN are from \citet{Dumouchel10}, for CO and C$^{18}$O are from \citet{Yang10a}, for N$_2$H$^+$ are from \citet{Daniel05a} and for NH$_3$ are from \citet{Danby88a} and \citet{Maret09a}.

Our procedure for modeling molecular line emission follows that of \citet{Onus18a}:
we first set the abundances of all species per H nucleus. The values we adopt are $X_{\rm HCN}=1.0\times10^{-8},X_{\rm CO}=1.0\times10^{-4},X_{{\rm C}^{18}{\rm O}}=1.0\times10^{-7},X_{{\rm N}_2{\rm H}^+}=1.0\times10^{-10},X_{{\rm pNH}_3}=1.0\times10^{-8}$ (where pNH$_3$ indicates para-NH$_3$, the isomer that produces the (1,1) inversion transition); these values are taken from \citet{Krumholz14b} and \citet{Offner08a}. Second, we assume a constant gas temperature $T=10$ K  \citep{Onus18a}. Under these assumptions we use \textsc{despotic} to compute a table of the luminosity per H$_2$ molecule in each line as a function of density and velocity gradient (which determines the optical depth in the LVG approximation), in a table of values running from $10^0$ to 10$^{10}$ cm$^{-3}$ in 100 logarithmically-spaced steps in number density and 10$^{-3}$ to 10$^3$ km s$^{-1}$ pc$^{-1}$ in 75 logarithmically-spaced steps in velocity gradient. For each cell in the simulation we take the line-of -sight (LOS) velocity gradient smoothed over 8 cells, and use that plus the density to determine the line luminosity in that cell by linearly interpolating in the table. We then generate position-position-velocity (PPV) cubes for each line using the software package \textsc{yt} \citep{Turk11a}. Each PPV cube has a resolution of $256^2\times200$, with a velocity range from $-4$ km s$^{-1}$ to 4 km s$^{-1}$. The corresponding resolution of a single PPV voxel is $\approx 0.02\, \mathrm{pc}\, \times\, 0.02 \,\mathrm{pc}\, \times \, 0.04 \, \mathrm{km}\,\mathrm{s}^{-1}$. We generate PPV cubes along each of the cardinal axes for each simulation at times $t/t_{\rm ff}=0$, $0.1$, $0.3$, and $0.6$.

\begin{figure*}
  \centering
  \includegraphics[width=.8\linewidth]{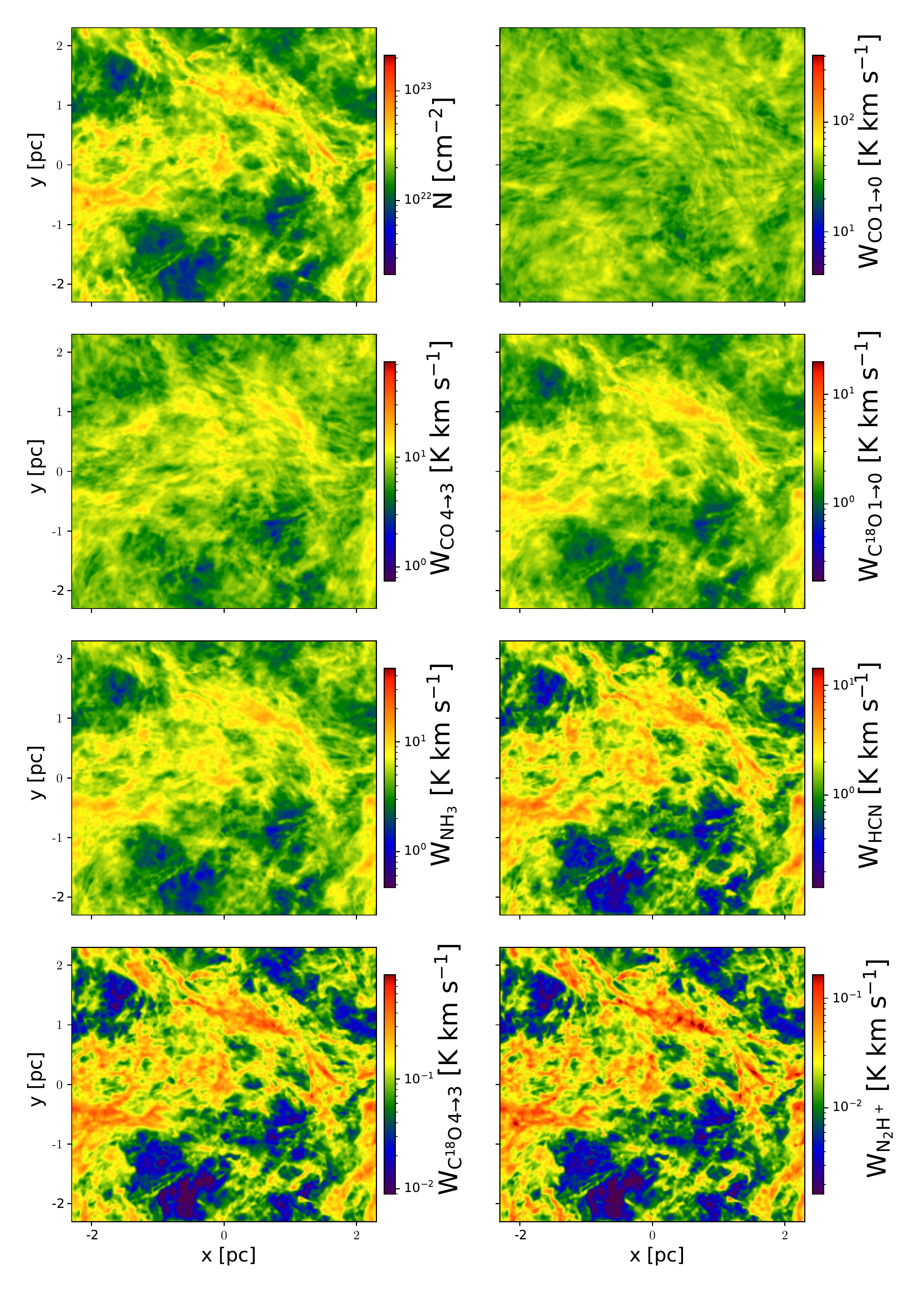}
\caption{Velocity-integrated intensity maps for seven different line tracers as indicated in each panel for the $\beta=0.2$ run at $t=0.1 t_{\mathrm{ff}}$. We also show the true column density map in the top left panel. The line of sight is perpendicular to the direction of the mean magnetic field. The colour bars in each panel have a dynamic range of 100, and are all centred on the mean pixel value, enabling direct comparisons between the panels. }
\label{fig:integrated_intensity_maps}
\end{figure*}

\subsection{Modeling real telescopic observations}
\label{ssec: model_telescope}

Real observational surveys always have finite signal-to-noise ratio (SNR) and finite spatial and spectral resolution. In order to compare our synthetic observations to real observations on an equal basis, we must therefore model these effects. For this purpose, we select resolutions and sensitivities typical of Galactic surveys, since the small size of our simulated region ($4.6$ pc) makes comparison to extragalactic studies problematic. We consider SNRs of 5, 10 and 20, a beam size of 0.1 pc, and a velocity channel width of 0.08 km s$^{-1}$. This spatial and spectral resolution is comparable to that of wide-area surveys such as COMPLETE \citep{Ridge06a} or the Green Bank Ammonia Survey \citep{Friesen17a}.

Our implementation of telescope effects is as follows: we first convolve the image in each PPV velocity channel with a Gaussian beam with a size of 0.1 pc to simulate the effect of beam-smearing. Second, we coarsen our original PPV cube to the target spatial and spectral resolution. Third, we add noise to the voxels in our PPV cube. The noise assigned for each voxel is drawn from a Gaussian distribution with a dispersion that is equal to the mean luminosity in the zero-velocity channel in the noise-free map, divided by the SNR. For the purpose of the analysis below, we mask all voxels in which the total signal, after noise is added, is below 3 times the noise level. Similarly, for velocity-integrated quantities, we mask pixels for which the intensity integrated along the line of sight is lower than $I_{\mathrm{noise}} = \sigma_{\mathrm{lum}}\Delta v \sqrt{n_{\mathrm{channel}}}$, where $\sigma_{\rm lum}$ is the noise level per channel, $\Delta v$ is the channel width, and $n_{\rm channel}$ is the number of channels in the image.

\begin{table*}

\caption{Summary of results}
\begin{center}
    \begin{tabular}{cc|c|c|cccccccc} \hline\hline
    \multicolumn{2}{c|}{Snapshot} & & & \multicolumn{5}{c}{Line} \\
    $\beta$ & $t/t_{\rm ff}$ & Quantity & True & CO $1\to0$ & CO $1\to0$ thin & CO $4\to3$ & C$^{18}$O $1\to0$ & NH$_3$ & HCN & C$^{18}$O $4\to3$ & N$_2$H$^+$ \\ \hline
    
    \multirow{ 5}{*}{0.2} & \multirow{5}{*}{0.1} & 
    $\log \langle n \rangle_{\rm L}$ [cm$^{-3}$] & 3.90 & 3.24 & & 3.49 & 3.76 & 3.75 & 3.96 & 4.10 & 4.21 \\
    & & $\langle\tau\rangle_{\rm M}$ & -- & 1630 & & 491 & 1.53 & 7.91 & 35.7 & 0.527 & 0.105 \\
    & & $\langle\sigma_{\parallel}\rangle$ [km s$^{-1}$] & 0.59 & 0.71 & 0.58 & 0.64 & 0.61 & 0.59 & 0.52 & 0.47 & 0.45 \\ 
    & & \multirow{2}{*}{$\langle \sigma_{\perp}\rangle$ [km s$^{-1}$]} & 0.58 & 0.73 & 0.57 & 0.64 & 0.60 & 0.58 & 0.52 & 0.47 & 0.45 \\
    & & & 0.56 & 0.68 & 0.56 & 0.61 & 0.58 & 0.57 & 0.52 & 0.48 & 0.45 \\
    \hline
     
    \multirow{5}{*}{0.2} & \multirow{5}{*}{0.3} & 
    $\log \langle n \rangle_{\rm L}$ [cm$^{-3}$] & 4.39 & 3.25 & & 3.52 & 3.85 & 3.81 & 4.05 & 4.26 & 4.49  \\
    & & $\langle\tau\rangle_{\rm M}$ & -- & 3500 & & 1060 & 3.27 & 16.7 & 64.8 & 1.16 & 0.155 \\
    & & $\langle\sigma_{\parallel}\rangle$ [km s$^{-1}$] & 0.57 & 0.69 & 0.57 & 0.63 & 0.60 & 0.58 & 0.51 & 0.45 & 0.41 \\ 
    & & \multirow{2}{*}{$\langle \sigma_{\perp}\rangle$ [km s$^{-1}$]} & 0.59 & 0.75 & 0.59 & 0.65 & 0.62 & 0.60 & 0.54 & 0.48 & 0.45 \\
    & & & 0.55 & 0.67 & 0.54 & 0.60 & 0.57 & 0.56 & 0.51 & 0.46 & 0.43 \\
    \hline
    
    \multirow{5}{*}{0.2} & \multirow{5}{*}{0.6} & 
    $\log \langle n \rangle_{\rm L}$ [cm$^{-3}$] & 5.42 & 3.27 & & 3.57 & 4.00 & 4.91 & 4.19 & 4.51 & 4.97 \\
    & & $\langle\tau\rangle_{\rm M}$ & -- & 37500 & & 11300 & 34.8 & 176 & 598 & 12.6 & 1.05 \\
    & & $\langle\sigma_{\parallel}\rangle$ [km s$^{-1}$] & 0.54 & 0.69 & 0.54 & 0.62 & 0.58 & 0.57 & 0.49 & 0.44 & 0.40 \\
    & & \multirow{2}{*}{$\langle \sigma_{\perp}\rangle$ [km s$^{-1}$])} & 0.61 & 0.77 & 0.60 & 0.68 & 0.65 & 0.63 & 0.56 & 0.50 & 0.45 \\
    & & & 0.53 & 0.66 & 0.53 & 0.60 & 0.57 & 0.56 & 0.51 & 0.45 & 0.41 \\
    \hline
    
    \multirow{5}{*}{2} & \multirow{5}{*}{0.1} & 
    $\log \langle n \rangle_{\rm L}$ [cm$^{-3}$] & 3.87 & 3.25 & & 3.48 & 3.73 & 3.73 & 3.93 & 4.07 & 4.18 \\
    & & $\langle\tau\rangle_{\rm M}$ & -- & 1670 & & 504 & 1.57 & 8.14 & 37.4 & 0.538 & 0.114 \\
    & & $\langle\sigma_{\parallel}\rangle$ [km s$^{-1}$] & 0.57 & 0.65 & 0.57 & 0.62 & 0.59 & 0.58 & 0.53 & 0.49 & 0.47 \\
    & & \multirow{2}{*}{$\langle \sigma_{\perp}\rangle$ [km s$^{-1}$]} & 0.51 & 0.57 & 0.51 & 0.55 & 0.53 & 0.52 & 0.49 & 0.45 & 0.43 \\
    & & & 0.54 & 0.61 & 0.54 & 0.58 & 0.56 & 0.55 & 0.51 & 0.48 & 0.47 \\
     \hline
    
    \multirow{5}{*}{2} & \multirow{5}{*}{0.3} & 
    $\log \langle n \rangle_{\rm L}$ [cm$^{-3}$] & 4.16 & 3.28 & & 3.54 & 3.85 & 3.82 & 4.05 & 4.25 & 4.45 \\
    & & $\langle\tau\rangle_{\rm M}$ & -- & 3700 & & 1120 & 3.46 & 17.6 & 68.7 & 1.23 & 0.166 \\
    & & $\langle\sigma_{\parallel}\rangle$ [km s$^{-1}$] & 0.58 & 0.66 & 0.57 & 0.63 & 0.60 & 0.59 & 0.53 & 0.48 & 0.45 \\
    & & \multirow{2}{*}{$\langle \sigma_{\perp}\rangle$ [km s$^{-1}$]} & 0.51 & 0.58 & 0.51 & 0.56 & 0.53 & 0.53 & 0.49 & 0.45 & 0.42 \\
    & & & 0.53 & 0.61 & 0.53 & 0.58 & 0.55 & 0.55 & 0.50 & 0.46 & 0.43 \\
     \hline
    
    \multirow{5}{*}{2} & \multirow{5}{*}{0.6} & 
    $\log \langle n \rangle_{\rm L}$ [cm$^{-3}$] & 5.73 & 3.34 & & 3.63 & 4.07 & 3.97 & 4.24 & 4.60 & 5.20 \\
    & & $\langle\tau\rangle_{\rm M}$ & -- & 67400 & & 20400 & 62.6 & 316 & 1070 & 22.7 & 1.85 \\
    & & $\langle\sigma_{\parallel}\rangle$ [km s$^{-1}$] & 0.58 & 0.69 & 0.58 & 0.64 & 0.61 & 0.60 & 0.54 & 0.49 & 0.45 \\
    & & \multirow{2}{*}{$\langle \sigma_{\perp}\rangle$ [km s$^{-1}$]} & 0.53 & 0.60 & 0.53 & 0.58 & 0.56 & 0.56 & 0.52 & 0.48 & 0.45 \\
    & & & 0.54 & 0.64 & 0.54 & 0.62 & 0.58 & 0.58 & 0.52 & 0.47 & 0.43 \\
     \hline
    
    \multirow{5}{*}{20} & \multirow{5}{*}{0.1} & 
    $\log \langle n \rangle_{\rm L}$ [cm$^{-3}$] & 3.95 & 3.32 & & 3.54 & 3.79 & 3.79 & 4.01 & 4.16 & 4.29 \\
    & & $\langle\tau\rangle_{\rm M}$ & -- & 1560 & & 473 & 1.47 & 7.65 & 35.2 & 0.505 & 0.105 \\
    & & $\langle\sigma_{\parallel}\rangle$ [km s$^{-1}$] & 0.67 & 0.70 & 0.66 & 0.69 & 0.68 & 0.67 & 0.63 & 0.59 & 0.57 \\
    & & \multirow{2}{*}{$\langle \sigma_{\perp}\rangle$ [km s$^{-1}$]} & 0.49 & 0.57 & 0.49 & 0.55 & 0.51 & 0.51 & 0.46 & 0.42 & 0.39 \\
    & & & 0.57 & 0.65 & 0.56 & 0.62 & 0.59 & 0.58 & 0.53 & 0.48 & 0.45 \\
     \hline
    
    \multirow{5}{*}{20} & \multirow{5}{*}{0.3} & 
    $\log \langle n \rangle_{\rm L}$ [cm$^{-3}$] & 5.23 & 3.40 & & 3.65 & 4.01 & 3.86 & 4.22 & 4.47 & 4.90 \\
    & & $\langle\tau\rangle_{\rm M}$ & -- & 5480 & & 1660 & 5.10 & 25.9 & 95.5 & 1.83 & 0.203 \\
    & & $\langle\sigma_{\parallel}\rangle$ [km s$^{-1}$] & 0.66 & 0.73 & 0.66 & 0.70 & 0.68 & 0.67 & 0.62 & 0.58 & 0.55 \\
    & & \multirow{2}{*}{$\langle \sigma_{\perp}\rangle$ [km s$^{-1}$]} & 0.48 & 0.58 & 0.48 & 0.55 & 0.51 & 0.51 & 0.47 & 0.42 & 0.39 \\
    & & & 0.58 & 0.67 & 0.58 & 0.65 & 0.61 & 0.60 & 0.55 & 0.50 & 0.46 \\
     \hline
    
    \multirow{5}{*}{20} & \multirow{5}{*}{0.6} & 
    $\log \langle n \rangle_{\rm L}$ [cm$^{-3}$] & 6.35 & 3.53 & & 4.85 & 4.41 & 4.25 & 4.53 & 5.00 & 5.73 \\
    & & $\langle\tau\rangle_{\rm M}$ & -- & 29600 & & 8950 & 27.5 & 139 & 473 & 9.95 & 0.834 \\
    & & $\langle\sigma_{\parallel}\rangle$ [km s$^{-1}$] & 0.65 & 0.75 & 0.65 & 0.68 & 0.67 & 0.64 & 0.58 & 0.56 & 0.54 \\
    & & \multirow{2}{*}{$\langle \sigma_{\perp}\rangle$ [km s$^{-1}$]} & 0.56 & 0.68 & 0.56 & 0.64 & 0.60 & 0.60 & 0.55 & 0.49 & 0.45 \\
    & & & 0.64 & 0.78 & 0.64 & 0.73 & 0.68 & 0.68 & 0.62 & 0.55 & 0.52 \\
    \hline\hline
    \end{tabular}
\end{center}
\label{tab:results_summary}
\end{table*}

\section{Results}
\label{sec: results}

In this section we mainly focus on the snapshots of $\beta=0.2$ and $t=0.1t_{\mathrm{ff}}$ and $t=0.6t_{\mathrm{ff}}$, using projections in which the orientation is perpendicular to the magnetic field. We discuss the dependence of the results over the full parameter space in \aref{appd:full_par}, where we show that our qualitative conclusions hold regardless of the snapshots we choose to analyse. For reasons of simplicity, we therefore focus on these two example cases in the main body of the paper. For the first part of this section we use our noise-free maps at the native resolution of the simulation; we defer discussion of the biases induced by noise and finite resolution to \autoref{ssec:telescope_effect}.

\subsection{Qualitative Results}
\label{ssec: qualitative results}

We show an example true column density map and integrated intensity maps for our seven different tracers for the case $\beta=0.2$, t=0.1t$_{\mathrm{ff}}$ (i.e. just after gravity is turned on) in \autoref{fig:integrated_intensity_maps}. In order to facilitate comparisons between different tracers, the dynamic range is the same in every panel. We see that different tracers pick up different parts of the flow, as expected \citep[e.g.,][]{Burkhart13a}. Due to strong optical depth effects, CO shows a smaller dynamic range in column density than is actually present, and preferentially picks out lower density regions. Conversely, dense gas tracers such as HCN, C$^{18}$O J=4$\to$3, and N$_2$H$^+$ produce emission primarily from overdense regions, and show much larger deficits along low column density lines of sight than are actually present. C$^{18}$O J=1$\to0$ and NH$_3$ sit in between these extremes, reproducing the dynamic range found in the true column density map relatively well.
\begin{figure}
\centering
  \includegraphics[width=1\columnwidth]{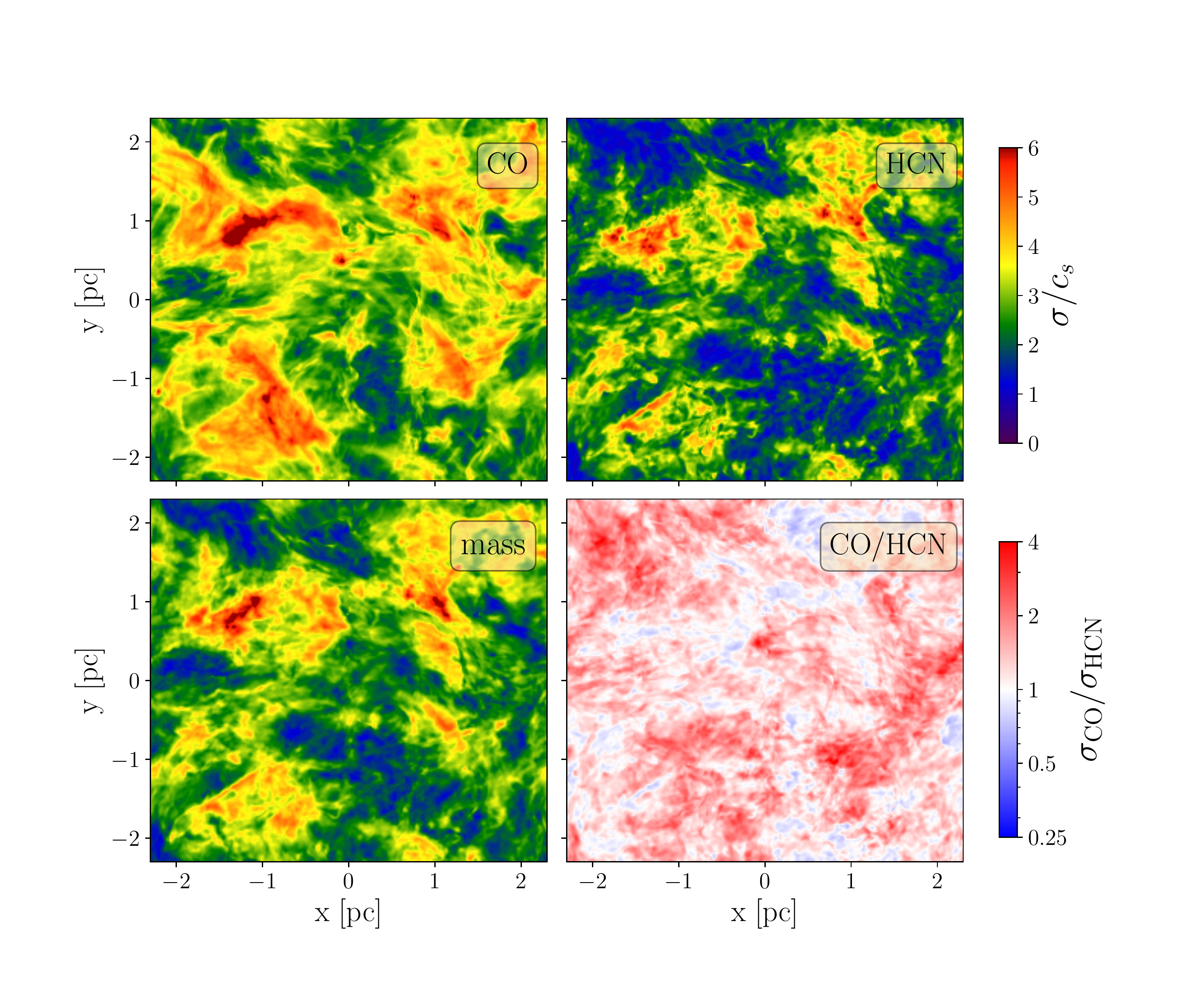}
\caption{Luminosity-weighted second moment maps for CO J = $1\to0$ (top left), HCN (top right), and for their ratio (bottom right) for the same snapshot and projection as shown in \autoref{fig:integrated_intensity_maps}. The bottom left panel shows the true, mass-weighted second moment map. In all cases the second moments we plot are normalised to the gas sound speed $c_s = 0.2$ km s$^{-1}$.
}
\label{fig:second_moment}
\end{figure}

In order to analyze the complex statistical properties of the velocity structure, in each pixel we calculate the luminosity-weighted first moment
\begin{equation}
    \bar{v} = \frac{\int L_v v \, dv}{L}
\end{equation}
and second moment
\begin{equation}
    \sigma_v =\left[\frac{\int L_v (v-\bar{v})^2\, dv}{L}\right]^{1/2},
\end{equation}
where $L_v$ is the specific luminosity at velocity $v$ and $L=\int L_v \, dv$ is the velocity-integrated luminosity. \autoref{fig:second_moment} shows example second moment maps for HCN and CO J=1$\to0$, as well as their ratio, for the same snapshot as shown in \autoref{fig:integrated_intensity_maps}.

We summarise the second moments that we measure for each simulation snapshot and each orientation in \autoref{tab:results_summary}. In this table, we report the luminosity-weighted mean second moment for each snapshot $\int L \sigma_v \, dA / \int L \, dA$, where the integral is over all pixels in the PPV cube. 
For comparison, we also calculate the true mass-weighted velocity dispersion $\int \Sigma \sigma_v\, dA / \int \Sigma\, dA$, where $\Sigma$ is the surface density of a pixel and $\sigma_v$ is the mass-weighted mean velocity dispersion along that pixel. This gives the velocity dispersion without bias from the density-dependence of emission tracers or optical depth effects. For both the true and measured second moments,
we distinguish between measurements in the direction parallel to the magnetic flux, which we denote $\langle \sigma_{\parallel}\rangle$, and measurements in the two directions perpendicular to the magnetic flux, which we denote $\langle \sigma_{\perp}\rangle$; since there are two cardinal directions perpendicular to the field, we list two values of $\langle \sigma_{\perp}\rangle$ in \autoref{tab:results_summary}.

In both the example shown in \autoref{fig:second_moment}, and in the numerical values reported in \autoref{tab:results_summary}, we see that our simulated maps exhibit the general trend that motivates much of this study: some tracers such as CO J=1$\to$0 show large, highly-supersonic second moments, while others such as NH$_3$ or N$_2$H$^+$ show systematically smaller second moments, which approach transsonic values in some cases. Which is closest to the true, mass-weighted velocity dispersion varies depending on the observation direction and the snapshot. In the remainder of this section, we investigate the physical reasons for these trends.

\subsection{Density effects}
\label{ssec:Line Width Size Relation vs. Density}

\begin{figure}
  \centering
  \includegraphics[width=\columnwidth]{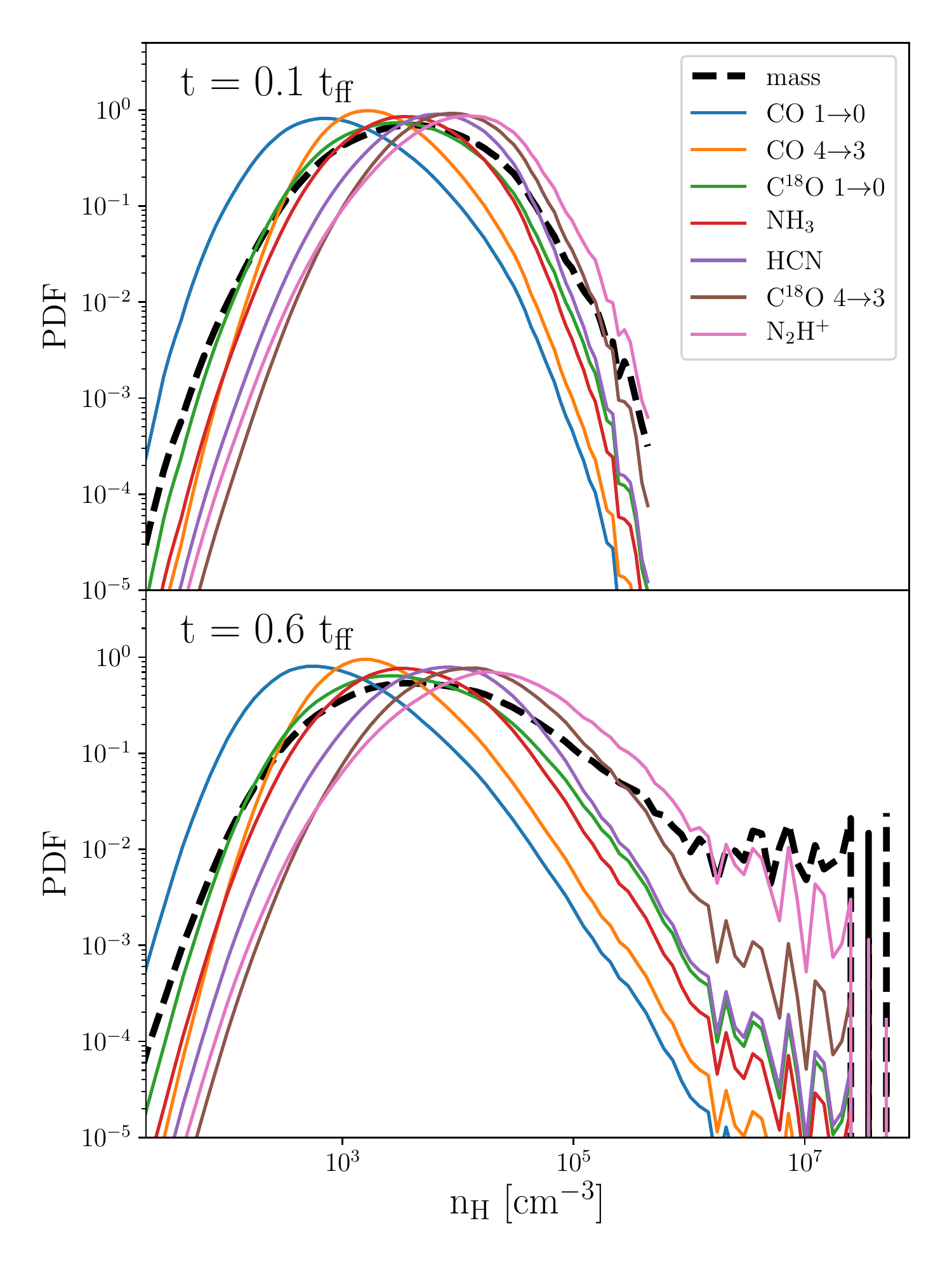}
\caption{
PDF of luminosity (coloured lines) and mass (black dashed line) as a function of log density for all the tracers for the same simulation as shown in \autoref{fig:integrated_intensity_maps} ($\beta=0.2$) at times $t=0.1t_{\rm ff}$ and $t=0.6t_{\rm ff}$, corresponding to states early and late in the star formation process.} 
\label{fig:L-rho}
\end{figure}

One obvious difference between molecular tracers is the densities of gas to which they are sensitive. We illustrate this in \autoref{fig:L-rho}, which shows the PDF of luminosity with respect to gas density for all the tracers and in the same simulation as shown in \autoref{fig:integrated_intensity_maps}, at two different times, one early in the evolution ($t=0.1t_{\rm ff}$) and one after the collapse is well-
advanced ($t=0.6t_{\rm ff}$). We can see that different tracers are sensitive to different ranges of density. Some, such as CO J=1$\to0$, yield a majority of their emission from gas that is less dense than the mass-weighted mean, while others, such as N$_2$H$^+$, are biased to gas that is denser than the mean; for this particular simulation, C$^{18}$O J=$1\to0$ NH$_3$ appear to be a reasonably good tracer of the true density structure, at least near the peak of the density PDF, though this is not true of all simulations at all times.

\begin{figure}
\centering
  \includegraphics[width=1.\linewidth]{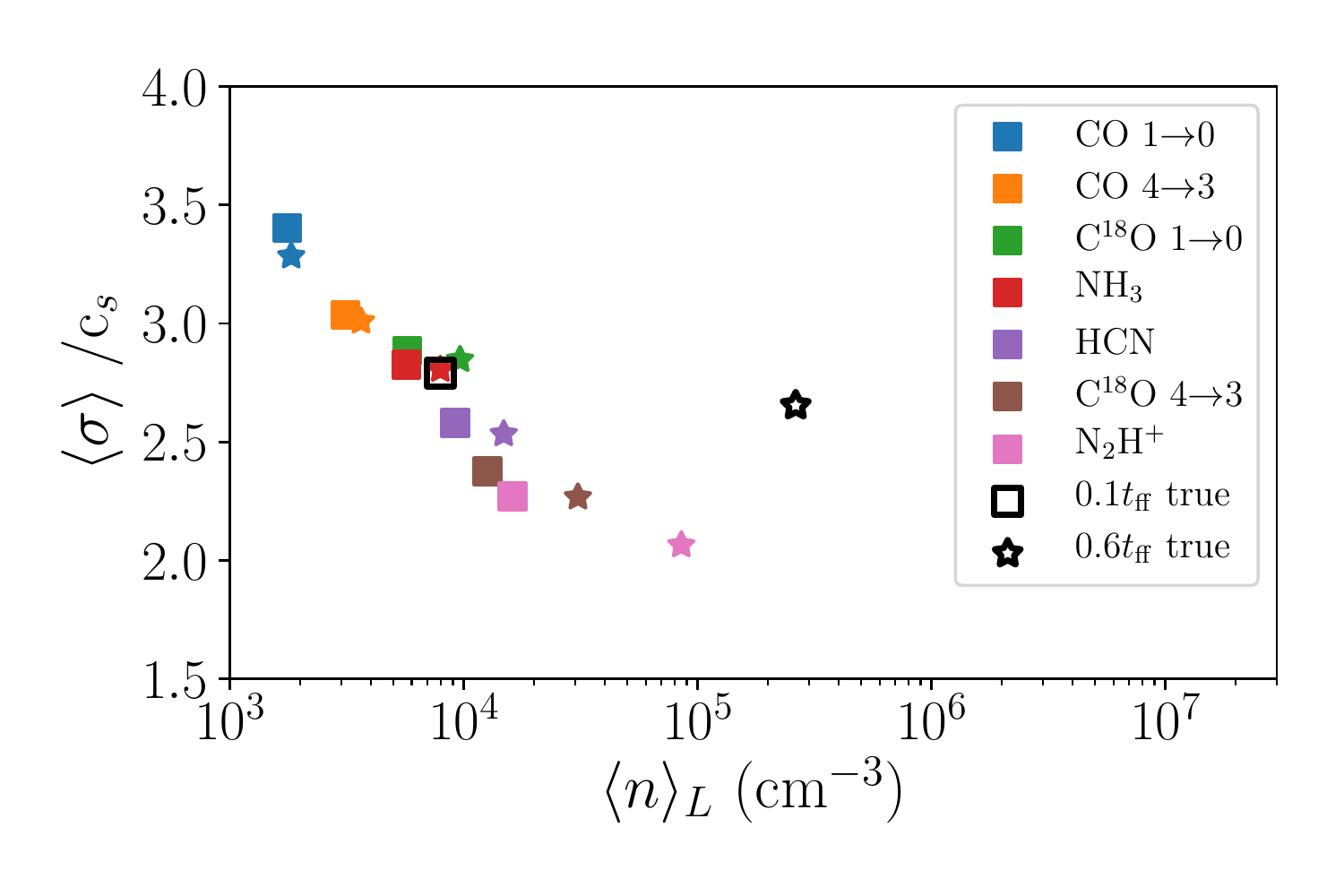}
\caption{Luminosity-weighted second moment vs.~luminosity-weighted mean density in the snapshots $\beta$=0.2, $t=0.1t_{\mathrm{ff}}$ and $t=0.6t_{\mathrm{ff}}$, projected along a direction perpendicular to the magnetic field; we show the data for other projections, times, and magnetic field strengths in Appendix A. Points are marker-coded by time and color-coded by different tracers. Open symbols, labelled ``true'' in the legend, show the true mass-weighted mean density and velocity dispersion for each snapshot.}
\label{fig:2nd_m_vs_Lrho}
\end{figure}

We investigate whether differences in linewidth are caused by density-dependent emission by comparing the mean second moments with the luminosity-weighted mean density. We define the latter quantity as
\begin{equation}
    \langle n \rangle_{\rm L} = \frac{\int n \mathcal{L}\, dV}{\int \mathcal{L} \, dV},
    \label{eq:n_L}
\end{equation}
where $\mathcal{L}$ is the luminosity per unit volume (integrated over all velocities) for a particular line and LOS as a function of position, $n$ is the number density (measured in terms of H nuclei per unit volume), and the integral runs over the entire simulation domain. We show the relationship between $\langle \sigma_\perp \rangle$ and $\langle n\rangle_{\rm L}$ for the snapshots of $\beta$=0.2 and $t=0.1t_{\mathrm{ff}}$ and $t=0.6t_{\mathrm{ff}}$ in \autoref{fig:2nd_m_vs_Lrho}, and report values of $\langle n\rangle_{\rm L}$ averaged over three cardinal axis for each snapshot in \autoref{tab:results_summary}. We also report values of the true mass-weighted mean density, which is simply given by \autoref{eq:n_L} with $\mathcal{L}$ set equal to the true density $\rho$. From the figure, we see that second moments are highly correlated with luminosity-weighted mean density. The velocity dispersion of the dense tracers can drop to trans-sonic values, despite the fact that the actual Mach number is 9, at least at early times. At later times the luminosity-weighted mean densities tend to increase, while the velocity dispersions remain roughly constant. This is a result of the decay of turbulence and the onset of collapse. However, even deep into the collapse, we see that velocity dispersion and luminosity-weighted mean density remain highly-correlated, and we therefore conclude that such correlations are a generic feature of turbulent flows, independent of whether they are self-gravitating or undergoing collapse. 

\begin{figure}
  \centering
  \includegraphics[width=\linewidth]{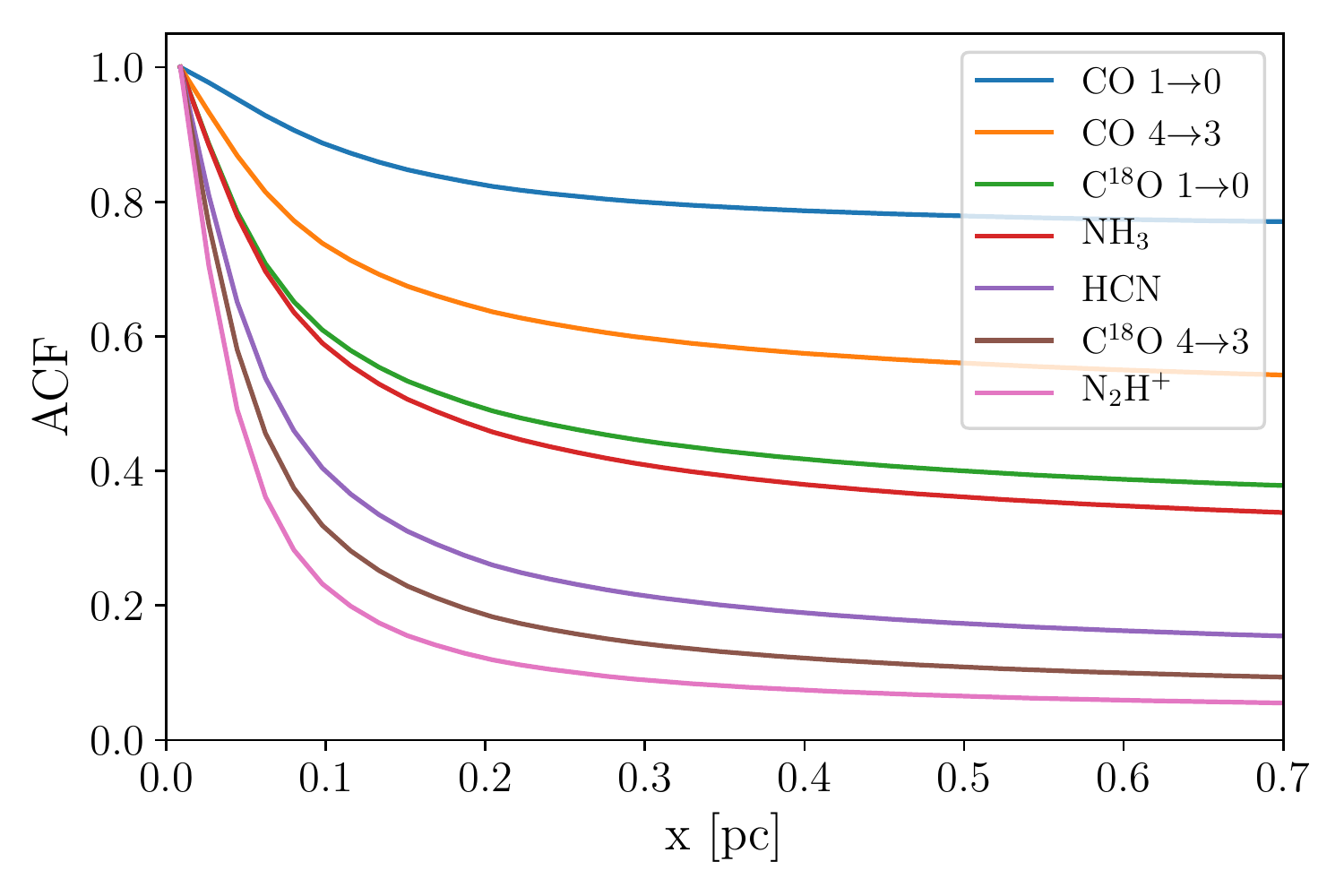}
\caption{1D auto-correlation function of the luminosity density for the same snapshot as shown in \autoref{fig:integrated_intensity_maps} for various tracers, as indicated in the legend.}
\label{fig:ACF}
\end{figure}

\begin{figure}
\centering
  \includegraphics[width=.8\linewidth]{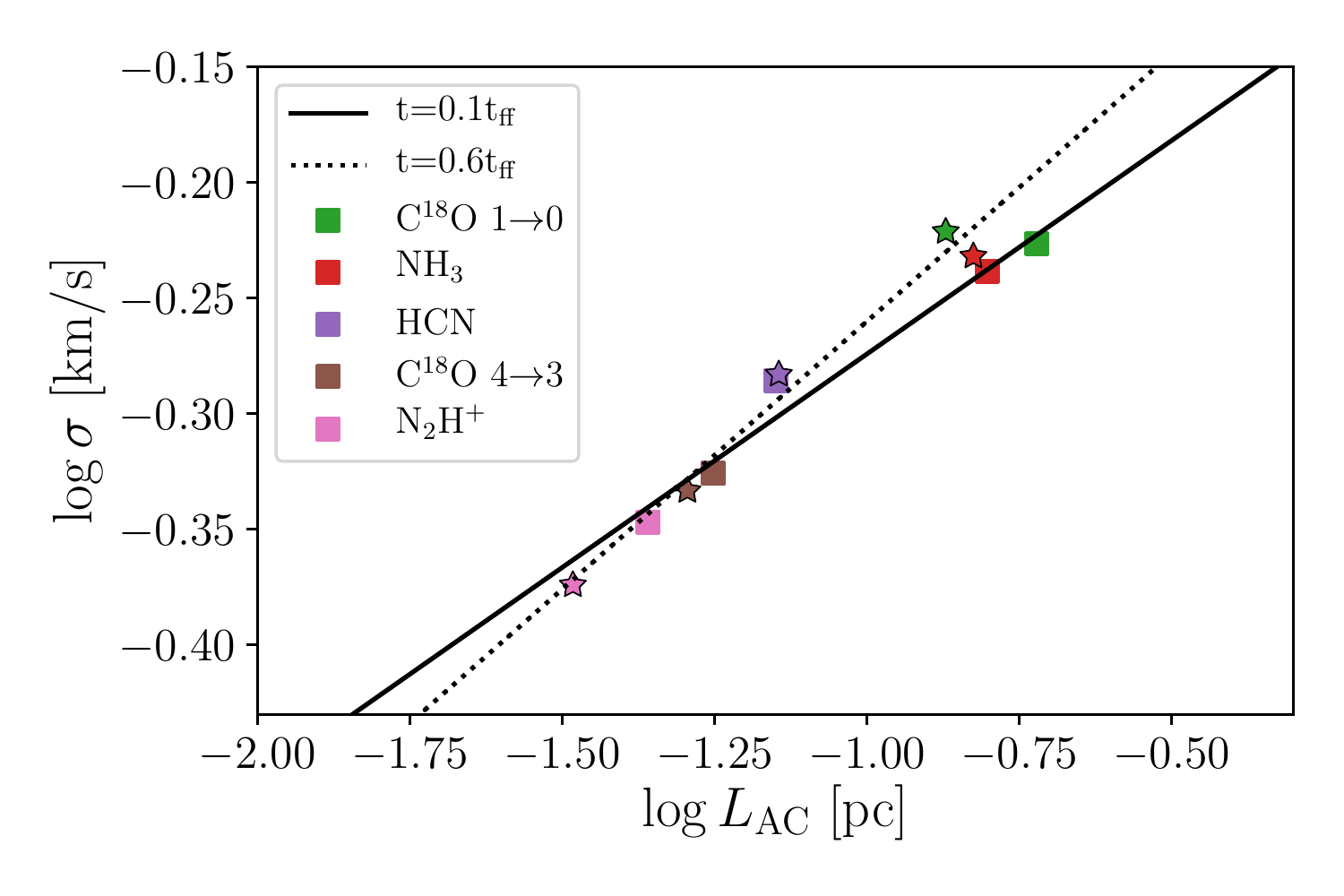}
\caption{Luminosity weighted second moment averaged over cardinal directions, $\sigma$, as a function of auto-correlation size of emitting regions $L_{\rm AC}$ for the snapshots $\beta$=0.2, $t=0.1t_{\mathrm{ff}}$ and $t=0.6t_{\mathrm{ff}}$. Different tracers as color coded in the same manner as \autoref{fig:2nd_m_vs_Lrho}. The solid and dotted lines represent least-squares linear fits to the data points for the corresponding time, as indicated in the legend. Note that this plot does not include CO J=1$\to0$ and CO J=4$\to3$, because we are unable to define $L_{\rm AC}$ for them.}
\label{fig:LS_relation}
\end{figure}

The results shown above strongly suggest different lines trace different regions, and this is at least partly drives the differences in linewidth. Such behaviour is generically expected in turbulent flows, which have power spectra $P(k)\propto k^\alpha$ with $\alpha < 0$, indicating that power resides predominantly on large scales. We can verify directly that this effect is at work by characterising the sizes of the emitting regions captured by different tracers, and checking how well these predict the velocity dispersion measured in that tracer. In order to characterise the sizes of the emitting regions, we calculate the auto-correlation function (ACF) of the luminosity density $\mathcal{L}$ for each tracer,
\begin{equation}
    A(\mathbf{x}) = \frac{\int \mathcal{L}(\mathbf{x}+\mathbf{x}')\mathcal{L}(\mathbf{x}') \, d^3x'}{\int \mathcal{L}(\mathbf{x}')\mathcal{L}(\mathbf{x}') \, d^3x'},
\end{equation}   
where $\mathbf{x}$ is known as the lag and the integral runs over the simulation volume. Note that we have not normalised the ACF by subtracting off the mean square of $\mathcal{L}$, because we are interested in the level of variation in the line compared to blank sky, not compared to the mean emission level of the cloud. Although our turbulence is not truly isotropic, due to the presence of a large-scale magnetic field, for convenience we will work with the angle-averaged 1D ACF, $A(x)$, which is simply the average of $A(\mathbf{x})$ over angle.
In \autoref{fig:ACF} we show the 1D ACF for the same snapshot as shown in \autoref{fig:integrated_intensity_maps}. We see that the ACF is different for different tracers, with low-density, high-optical depth tracers like CO J=1$\to0$ showing a shallow ACF, and high-density, low-optical depth tracers like N$_2$H$^+$ showing a steep ACF. For the purposes of our analysis here, we will define the characteristic auto-correlation length scale $L_{\rm AC}$ for a given tracer as the lag for which $A(L_{\rm AC}) = 0.5$. Note that this leaves $L_{\rm AC}$ undefined for CO J=1$\to0$ and J=4$\to$3, since the ACF for them remains above 0.5 even for lags comparable to the size of the simulation box.

We compare the measured linewidth in each tracer with the corresponding characteristic emitting size in \autoref{fig:LS_relation}. There is clearly a near-linear correlation between $\log L_{\rm AC}$ and $\log\sigma$, where $\sigma$ is the root mean square of the mean second moments measured along each of the three cardinal axes.
We illustrate this by plotting simple linear least-squares fits to the data in \autoref{fig:LS_relation}; these fits describe the data quite well, particularly at $t=0.1t_{\rm ff}$. Clearly at least part of the variation in linewidth measured with different tracers is a result of differences in density sensitivity leading to tracers picking out regions of different size. This, combined with the linewidth-size relation of turbulence, in turn induces a difference in linewidth between the tracers.

\subsection{Opacity effects}
\label{ssec:Opacity_Broadening}

\begin{figure}
  \centering
  \includegraphics[width=.8\linewidth]{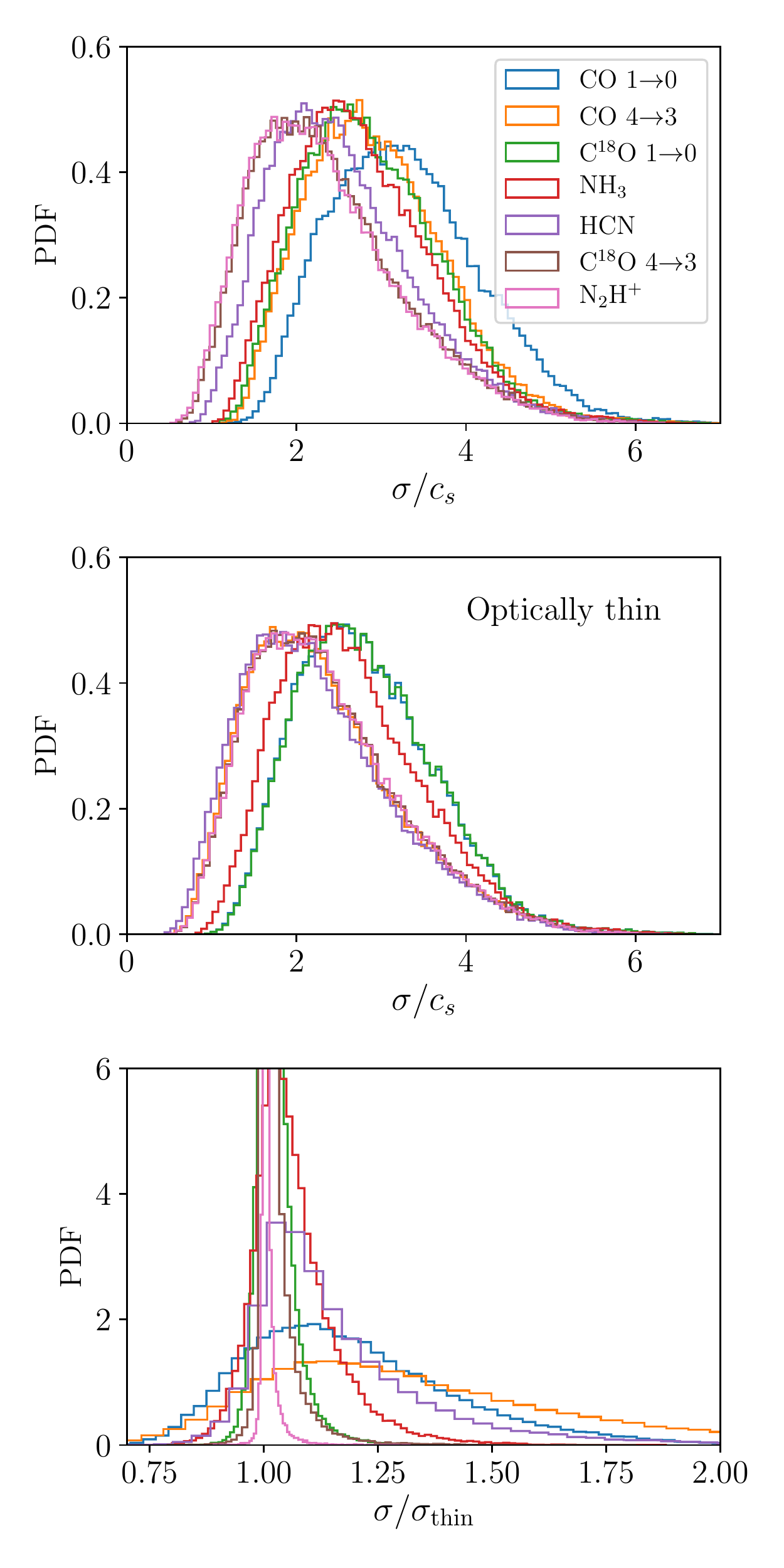}
\caption{PDF of luminosity-weighted second moment over all pixels for the same snapshot as \autoref{fig:integrated_intensity_maps}, measured along the line of sight perpendicular to the mean magnetic field. In the top panel we show the distribution of second moments measured in each pixel using our simulated line emission, including optical depth effects in the LVG approximation. In the middle panel, we show the second moments measured from line emission where we have artificially set the optical depths of all lines in all cells to zero. The bottom panel shows the distribution of the ratios of second moments computed including and ignoring optical depth effects. 
}
\label{fig:Opacity}
\end{figure}

We next explore the effects of opacity on the linewidths measured with optically thick tracers. As pointed out by \citet{Correia14a}, linewidths can be artificially enhanced by opacity broadening, whereby high optical depth suppresses emission in the line core more than in the line wings, making the line appear too broad. To begin exploring this effect, we use the cell-by-cell optical depths (which we compute using the LVG approximation) to calculate the mass-weighted mean optical depth $\langle\tau\rangle_{\rm M}$ for each of our simulation snapshots and LOS in each of our lines. We report these values averaged over three cardinal axis in \autoref{tab:results_summary}. As expected, we find that CO J $=1\to 0$ and CO J $=4\to 3$ are generally very optically thick ($\langle\tau\rangle_{\rm M}\sim 1000-10000$), HCN and CO J $=4\to 3$ are moderately optically thick ($\langle\tau\rangle_{\rm M}\sim 100-1000$), and all other lines are moderately or completely optically thin.

To see how this affects the inferred velocity dispersion, in the top panel of \autoref{fig:Opacity} we show the distribution of second moments of our seven tracers measured in every pixel for the same sample snapshot as shown in \autoref{fig:integrated_intensity_maps}. For comparison, in the middle panel we show the same quantity, but calculated in a case where we artificially set the optical depth of all lines to zero (or equivalently, where we take the limit of $\nabla\cdot\mathbf{v}\to\infty$ in the LVG approximation). In the bottom panel of the figure, we show the distribution of ratios of the measured to optically thin second moments; that is, the bottom panel is the distribution of the ratios of observed second moments including opacity effects (as shown in the top panel) to second moments that would be observed without opacity effects (as shown in the middle panel). From \autoref{fig:Opacity}, we see that opacity broadening is moderately strong for CO J$=1\to 0$, J$=4\to 3$ and HCN, on average adding $\sim 30\%$ to the CO-inferred velocity dispersion, $\sim 15\%$ to the HCN-inferred one. The effect is weak for all other lines. 

\begin{figure}
  \centering
  \includegraphics[width=\linewidth]{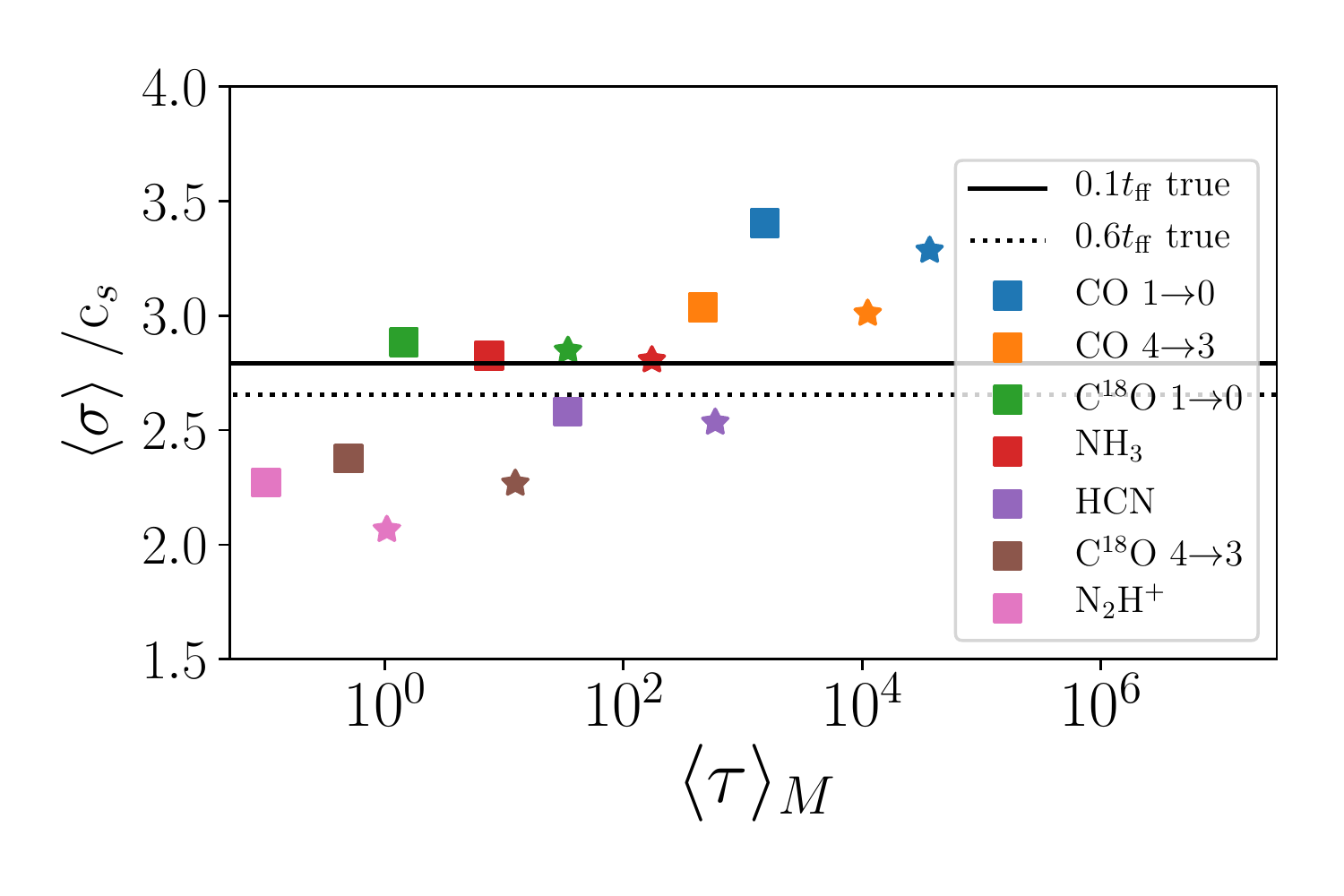}
\caption{Luminosity-weighted mean second moment versus mass-weighted mean opacity for $\beta=0.2$. Horizontal lines show the value of the true velocity dispersion at each time.}
\label{fig:second_moment-opacity}
\end{figure}

We investigate the dependence of linewidth on opacity for the snapshots of $\beta=0.2$ and $t=0.1t_{\mathrm{ff}}$ and $t=0.6t_{\mathrm{ff}}$ in \autoref{fig:second_moment-opacity}. From the figure, we see that there is a weak correlation between linewidth and opacity, consistent with our earlier finding that opacity broadening is a $\sim 30\%$ effect for CO J $=1\to0$ \citep{Correia14a} and a $\sim 15\%$ effect for HCN.  Interestingly, however, there is even a correlation between linewidth and opacity for mass-weighted mean opacities $\langle\tau\rangle_{\rm M} \lesssim 1$, where optical depth effects cannot possibly be important -- for example, \autoref{fig:Opacity} shows that optical depth effects are completely negligible for C$^{18}$O J $=1\to0$, J $=4\to3$ and N$_2$H$^+$, our three most optically thin-tracers, but there is nonetheless a systematic trend that linewidths measured with C$^{18}$O are larger than those measured with N$_2$H$^+$. 

The reason is simple: optical depth is correlated with density sensitivity, which we have also seen affects measured linewidths. Thus even in cases where the optical depth itself has no effect, there can still be an apparent correlation between optical depth and linewidth simply because the density range to which a given molecule is sensitive affects the linewidth, and density and optical depth are correlated. The relationship is even more complex for tracers that are at least marginally optically thick, because the effective critical density for a given species depends on its optical depth -- the level populations will thermalise in an optically thick region at lower density than in an optically thin one. Thus high optical depth weights the emission to lower density regions both because it suppresses the escape of photons from higher density ones, and because it helps thermalise the population and thus increase the luminosity in lower density ones.

In order to disentangle the various effects that optical depth has on line shape, we carry out the following experiment for CO. We first calculate the level population of CO using our normal escape probability treatment of optical depth effects, but we then calculate the resulting emission assuming the gas is optically thin. In this way we can separate out the effects of CO optical depth on the level population from its effects on the emergent light, i.e., the effects of opacity broadening. We calculate the velocity dispersion of the PPV cubes produced in this manner using the same procedure as in \autoref{ssec: qualitative results} and show the results in \autoref{tab:results_summary}. We see that the velocity dispersions computed for CO in this manner are generally very close to the values found for C$^{18}$O. This means that, at least for CO, the effect of opacity broadening is more important than the density sensitivity in setting the linewidth -- i.e., when we compute the density-dependence of emission including optical depth effects, but ignore the radiative transfer effects of optical depth, the linewidths we obtain are closer to the case where the optical depth is negligible for all purposes (as is the case for C$^{18}$O) than to the case where we include both optical depth effects in both the level population and the radiative transfer calculation. 
Conversely, for HCN, which has a more moderate optical depth and a stronger dependence on density, opacity broadening is clearly less important than density effects: while \autoref{fig:Opacity} indicates that opacity broadening does increase its linewidth, examination of \autoref{tab:results_summary} shows that it nonetheless yields a linewidth that is systematically \textit{smaller} than the true one. For HCN, density dependence is clearly more important than opacity dependence. 

Taken together, our experiments suggest that both density-dependent excitation and opacity broadening can have significant effects on inferred linewidths. For very optically thick species like CO $1 \to 0$, the opacity broadening effect is dominant. However, density-dependent excitation and the resulting variation in the characteristic sizes of emitting regions also produces a strong correlation between linewidth and the characteristic density of the emitting material. This primary correlation can also produce a spurious secondary correlation between optical depth and inferred linewidth even in species for which opacity broadening is completely negligible.

\subsection{Effects of finite resolution, sensitivity and beam-smearing}
\label{ssec:telescope_effect}

\begin{figure}
  \centering
  \includegraphics[width=1.1\columnwidth]{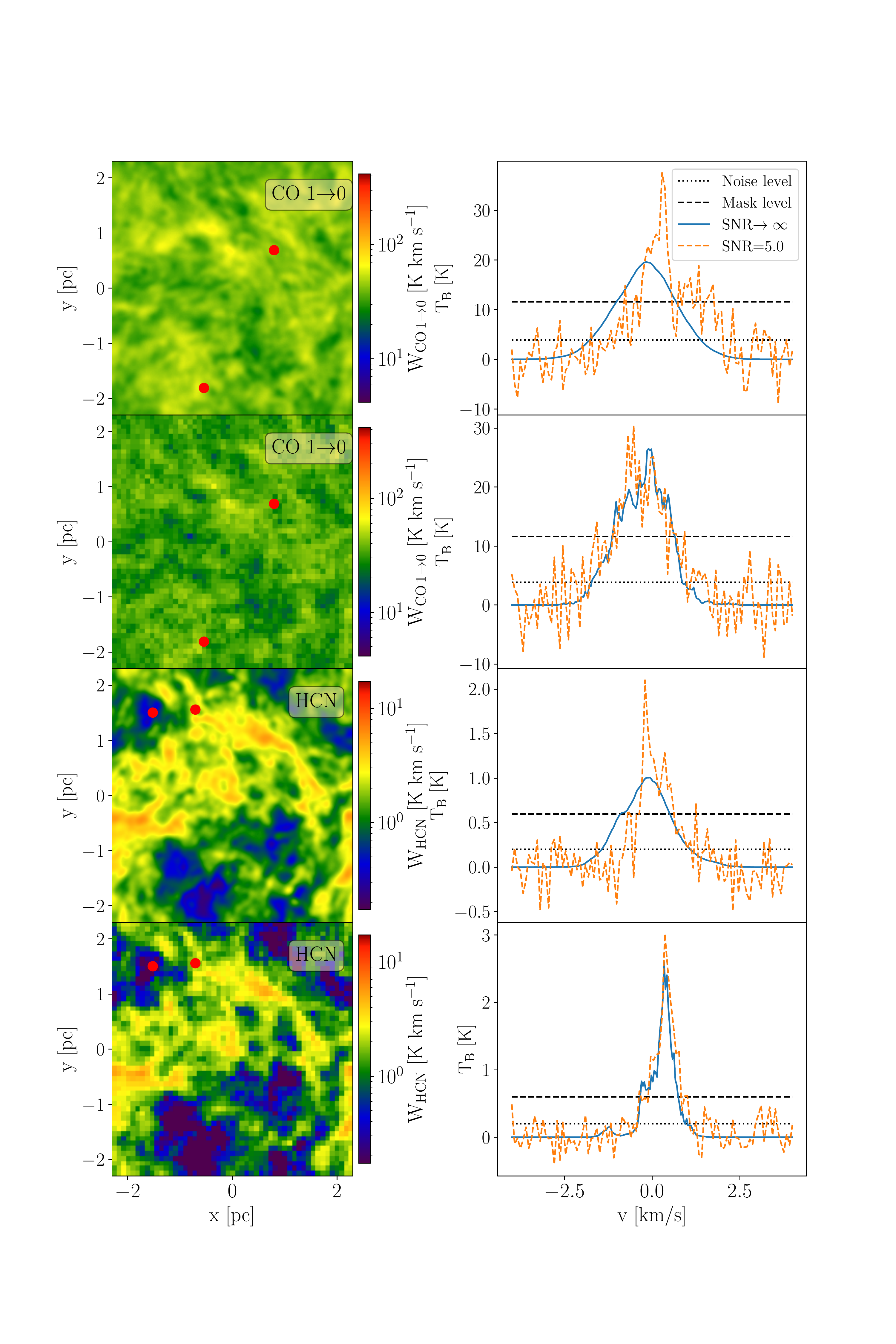}
\caption{Velocity-integrated intensity maps (left) and two typical spectra (right) for two tracers CO $1\to 0$ and HCN, for the same snapshot as \autoref{fig:integrated_intensity_maps}. In the left column, the top panel for each of the two tracers shows the results for the true PPV cube, while the bottom panel shows the results after beam convolution and with a finite SNR of 5. The red circles in the map denote the positions at which we extract the two example spectra shown in the right column. In this column, we show both the true and noise-added spectra; we also show the $1\sigma$ noise level and the mask we apply at $3\sigma$ as dotted and dashed lines, respectively.}
\label{fig:img&spct_noise}
\end{figure}

Finally, we investigate bias due to noise and beam-smearing. In \autoref{fig:img&spct_noise}, we show some examples of velocity-integrated intensity maps and typical spectra before and after adding noise.\footnote{Careful readers may notice that the brightness temperature in the CO lines is somewhat larger than the gas kinetic temperature of 10 K. While this should not happen in reality, it can happen in our simulations due to the limitations of the LVG approximation for radiative transfer, which treats all absorption as local, and thus can miss absorption of background emission by foreground structures that are located some distance from the emitter, but happen to overlap in velocity. This issue only affects CO, since no other tracer is optically-thick enough for spatially-distant foreground absorption to be significant. Moreover, by varying our method for approximating the velocity gradient, we have verified that this issue has no significant impact on our results for CO kinematics; changing our method of estimating of the velocity gradient such that the peak brightness temperature for CO changes by factors of $\sim 10$ produces $\lesssim 10\%$ changes in the inferred velocity dispersions.} In the left panel of \autoref{fig:img&spct_noise}, we see that the intensity maps are only minimally affected by noise and finite spatial resolution. However, in the right panel of \autoref{fig:img&spct_noise}, we see that for CO $1\to0$, the line wings are significantly hidden by the noise, which lowers the recovered linewidth, while for HCN this effect is much smaller. 

\begin{figure}
  \centering
  \includegraphics[width=\linewidth]{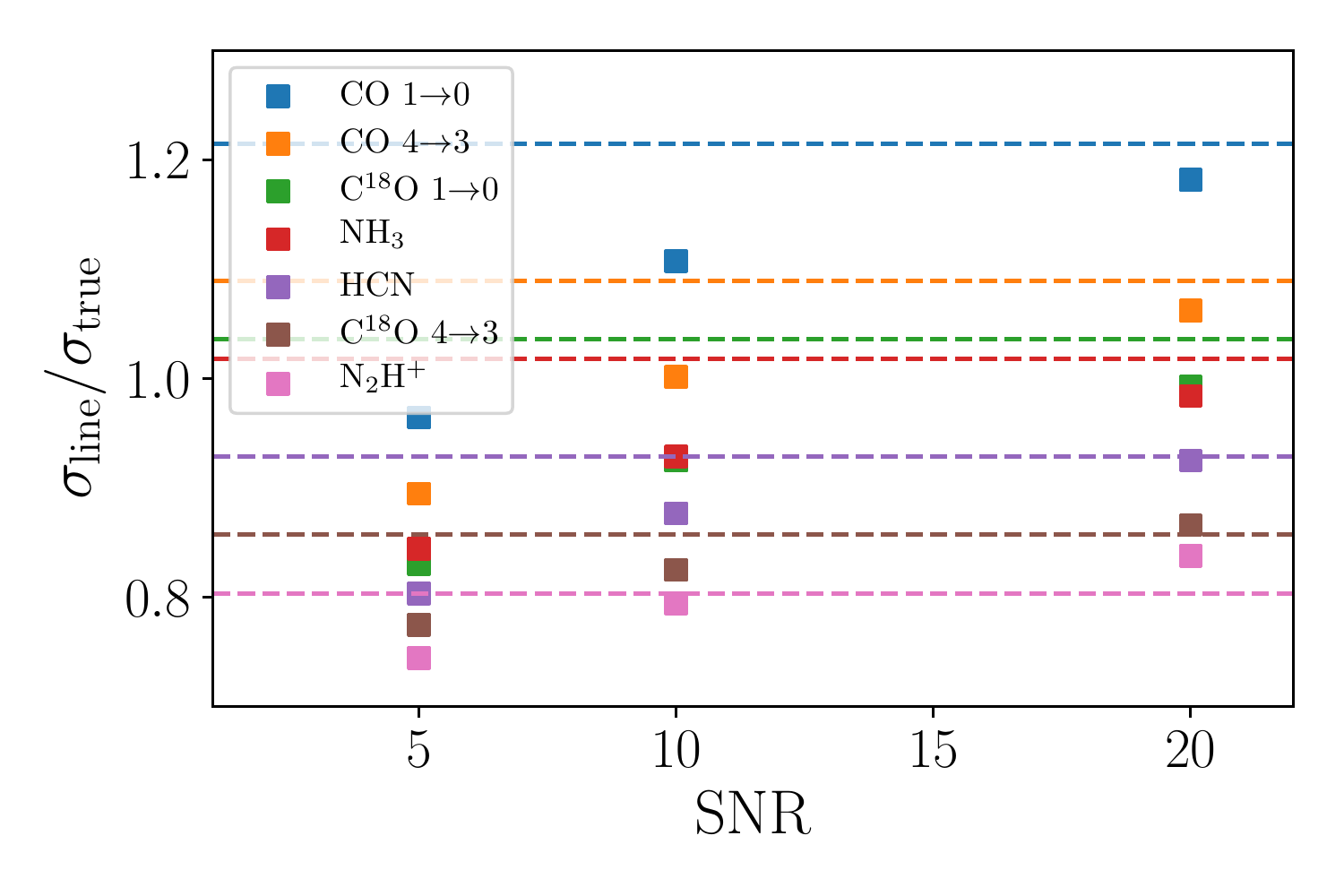}
\caption{
Ratio of the luminosity-weighted mean to true velocity dispersion, using dispersions inferred from seven emission lines as indicated in the legend, calculated from PPV cubes with finite resolution and added noise, as described in \autoref{ssec: model_telescope}. We show the results as a function of the signal-to-noise ratio. All points shown are for the same snapshot and projection as shown in \autoref{fig:integrated_intensity_maps}. For comparison, results obtained from the idealised synthetic observation (infinite SNR and resolution equal to the native resolution of the underlying simulations) are shown as dashed lines. }
\label{fig:teles_effect}
\end{figure}

In order to illustrate the dependence of this narrowing effect on the noise level and choice of tracer, in \autoref{fig:teles_effect} we show the ratio of the luminosity-weighted mean velocity dispersion inferred from our cubes with finite resolution and sensitivity to the true velocity dispersion. We show this ratio as a function of the signal-to-noise ratio of the observations. For comparison, we also show results obtained from the idealized synthetic observation (infinite SNR and high resolution) as the dashed lines. We see that limited SNR can lower the inferred linewidth significantly, especially for SNR of 5 and for low density tracers. This is because we throw out the portion of line wings contaminated by noise. In this sense, noise is the opposite of opacity effects -- the latter preferentially suppress the line centre, while the former suppresses the wings. At SNR $\sim$ 5, the linewidth we recover for CO $1 \to 0$ drops by $\sim$30\% compared to what we obtain in the infinite resolution limit, and even at SNR $\sim$ 20, it is still lowered by 5\%. For the highest density tracers such as N$_2$H$^+$, the bias induced by noise is smaller than for the low density tracers; for example, the N$_2$H$^+$ velocity dispersion we recover from the noisy cube is only 7\% smaller than for the true cube, even at a SNR of 5. Interestingly, at high SNR $\sim$ 20, the velocity dispersion inferred from the noisy cube can even slightly exceed the value recovered from the true cube, due to the effect of beam convolution. We have verified that this is the case by also constructing PPV cubes with beam smearing but no noise -- for such cubes, we find that the linewidths of the higher density tracers typically increase by a few percent, while those of the lower density tracers are largely unaffected.

To summarize, it seems that the bias introduced by telescope is set by a competition between beam and noise effects, and the bias induced by these two components is different for different tracers. Low density tracers are influenced significantly by noise and not affected much by beam-smearing, leading to lower measured velocity dispersions, whereas high density tracers are influenced less by noise and more by beam-smearing, so that the velocity dispersion we infer for them is increased. All of these effects of resolution and sensitivity sit on top of the radiative transfer and excitation effects we have explored in the previous sections.

\section{Discussion}
\label{sec: disc}

Having analysed the mechanisms that give rise to various biases, we are now in a position to draw overall conclusions about the relative reliability of various tracers, and how this depends on cloud properties. Doing so is our primary focus in this section.

\subsection{Which tracers reflect the true velocity dispersion?}
\label{ssec: Which tracers reflect the true velocity dispersion?}

\begin{figure}
  \centering
  \includegraphics[width=\linewidth]{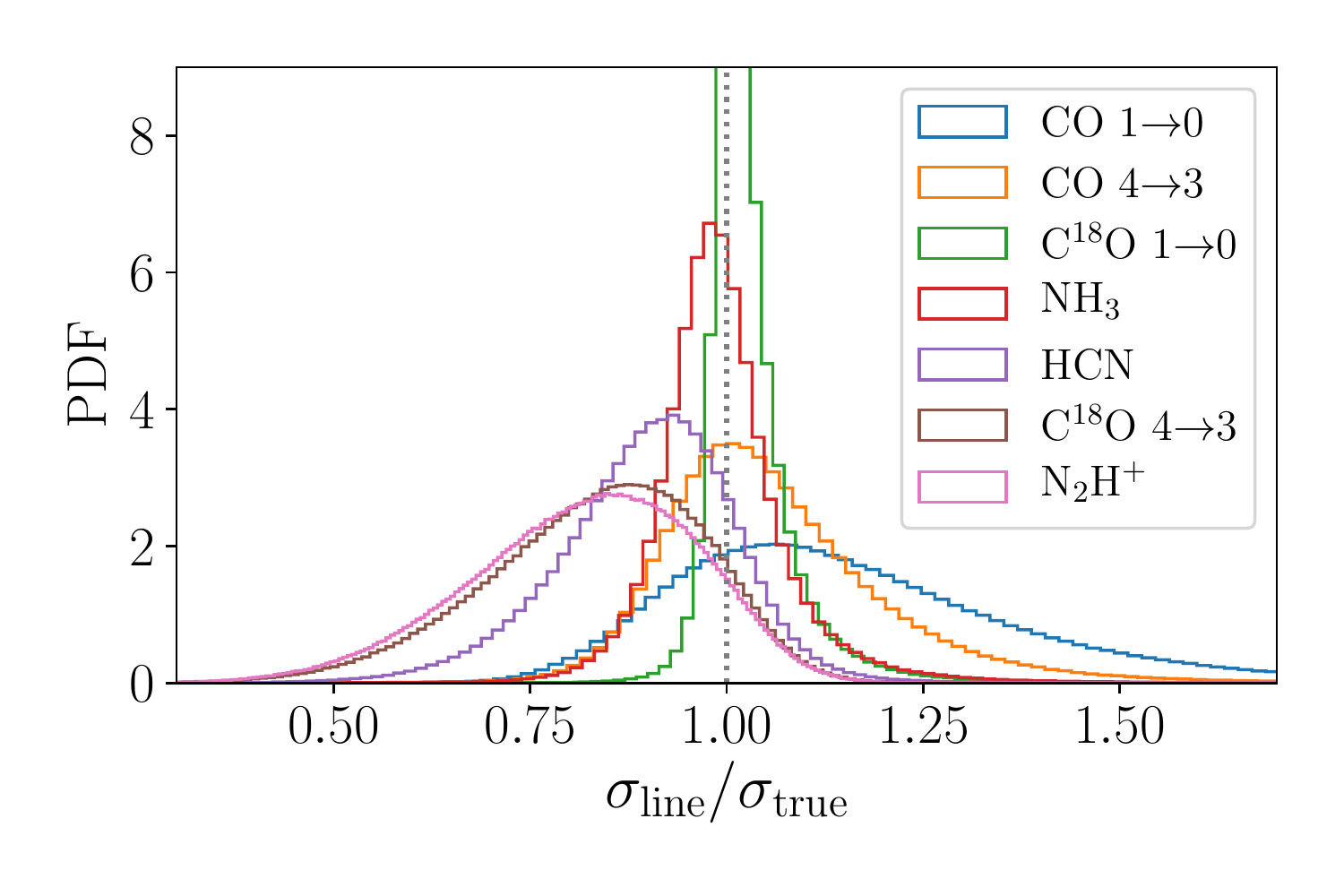}
\caption{
Distribution of ratio of the velocity dispersion measured using various emission lines, $\sigma_{\rm line}$, to the true mass-weighted velocity dispersion, $\sigma_{\rm true}$, for all pixels of all snapshots, i.e., combining all magnetic field strengths, times, and orientations. Different colours show different emission lines, as indicated in the legend.}
\label{fig:sigma_line_vs_sigma_true_all}
\end{figure}

We begin with the most basic question: which tracers most reliably match the true (i.e., mass-weighted) velocity dispersion, and what sorts of errors and biases do these and other tracers have? To answer this question, we plot the distribution of ratio of the velocity dispersion of emission lines to the true ones for all the pixels of all snapshots in \autoref{fig:sigma_line_vs_sigma_true_all}. We start here with the case without beam smearing or noise, and note that this histogram includes all snapshots at all times, not just the cases on which we focused as examples in \autoref{sec: results}. From this figure it is clear that, overall, C$^{18}$O is generally most accurate, with NH$_3$ as a close second; both have typical errors below $\sim 10\%$, and little bias, i.e., the PDF is reasonably well-centred around $\sigma_{\rm line}/\sigma_{\rm true} = 1$. Interestingly, we see that CO $4\to3$ is also well-centred on the true value. However, its distribution is significantly broader, with errors of $\sim 20\%$. This is not surprising, since we have seen that the good average performance of CO $4\to3$ is due to near-cancellation between density bias and opacity bias; the latter causes pixels with high column density to show inflated linewidths, while the former causes pixels with low column density to return linewidths that are artificially low.  CO $1\to0$ is biased high by $\sim 20\%$, and has a tail extending to $>50\%$, while the denser tracers C$^{18}$O $4\to3$ and N$_2$H$^+$ are biased low by a similar amount, and have tails extending down to a factor of 2 error.

We show in \aref{appd:full_par} that these general conclusions apply not just to the total distribution over all snapshots, but also to individual cases at different plasma $\beta$, orientation with respect to the magnetic field, and simulation time. It is at least somewhat surprising that which tracers are most accurate does not depend on these factors in light of \autoref{fig:L-rho}, which shows that which lines trace the \textit{mass} best does depend on evolutionary state -- at early times when the density distribution is close to lognormal, C$^{18}$O emission follows mass most closely, but at later times when the density distribution has developed a significant powerlaw tail, dense gas tracers such as N$_2$H$^+$ more accurately follow the tail of the density distribution.

The resolution to this apparent paradox can be found by noticing that, even at late times, C$^{18}$O remains the best tracer near the peak of the PDF. We have seen that density and velocity are anti-correlated, which is why dense gas tracers tend to be biased low in their estimates of the velocity dispersion. This effect helps protect the accuracy of moderate density tracers like C$^{18}$O and NH$_3$ at late times. Although there is substantial mass in the high-density powerlaw portion of the PDF, the bulk of the kinetic energy is still retained in the lower-density material for which C$^{18}$O and NH$_3$ remain accurate tracers. Thus the material that these tracers are failing to capture makes relatively little contribution to the velocity dispersion, and thus a failure to capture it introduces little error.

Finally, let us consider how beam smearing and noise change our picture as outlined above. From the analysis in \autoref{ssec:telescope_effect}, we see that SNR values as low as 5 will lead to measurements of the velocity dispersion that are up to $\sim 30\%$ lower than would be recovered in the limit of infinite SNR. High density tracers are the least affected, and become nearly insensitive to SNR once the SNR exceeds $\sim 10$, while low and moderate density tracers often require SNR of about 20 to approach the infinite SNR limit. Such high SNRs are generally only practical to obtain for the rotational lines of CO. This presents a challenge to observational survey design, because it is precisely such lines that suffer the most from opacity bias, and thus tend to \textit{overestimate} the velocity dispersion when the SNR is high. Conversely, observations of tracers such as C$^{18}$O and NH$_3$ that are relatively immune to density and opacity bias may often require long integration tines to reach acceptable SNR. In practice these considerations may suggest the use of CO $4\to3$ as the best available compromise, as it is the only line that gives a relatively precise measurement of kinematics but is also bright enough to allow reasonable mapping speeds at high SNR.

\subsection{Dependence on cloud density}
\label{ssec:density_dependence}

As discussed in \autoref{ssec:simulations}, in order to calculate observable line emission, we must choose a particular set of physical units for our simulation suite. It is therefore important to check to what extent our results are robust against this choice. In order to investigate this, we can rescale the simulations to arbitrary density and size scale. Since we are extracting an idealised sub-region of a molecular cloud, we are free to regard out simulation as representing a small, dense part of the cloud, or a larger, less dense part. Quantitatively, we rescale our density field by a factor $a$ compared to our fiducial choice, which means the average density becomes $n=1000a$ cm$^{-3}$. In the process, we have to fix the virial parameter, the Mach number, and the plasma beta, because these are all dimensionless quantities that affect the solutions to the equations of hydrodynamics. We also keep the sound speed the same, because that is set by the rate of cosmic ray heating, which is roughly constant in the Galaxy. In order to satisfy these constraints, we adopt following scalings for our re-scaled simulation:
\begin{eqnarray} 
t_{\rm ff}& = &1.1\cdot a^{-\frac{1}{2}}\;\mbox{Myr}\\
L_0& =& 4.6\cdot a^{-\frac{1}{2}}\;\mbox{pc}\\
v_{\rm rms}& = &1.8\;\mbox{km s}^{-1},\\
M &= &5900\cdot a^{-\frac{1}{2}} \;M_\odot\\
B_0& =& (13, 4.4, 1.3)\cdot a^{\frac{1}{2}}\;\mu\mbox{G}.
\end{eqnarray}
With these choices, all dimensionless numbers describing the flow are left unchanged.

\begin{figure}
  \centering
  \includegraphics[width=\linewidth]{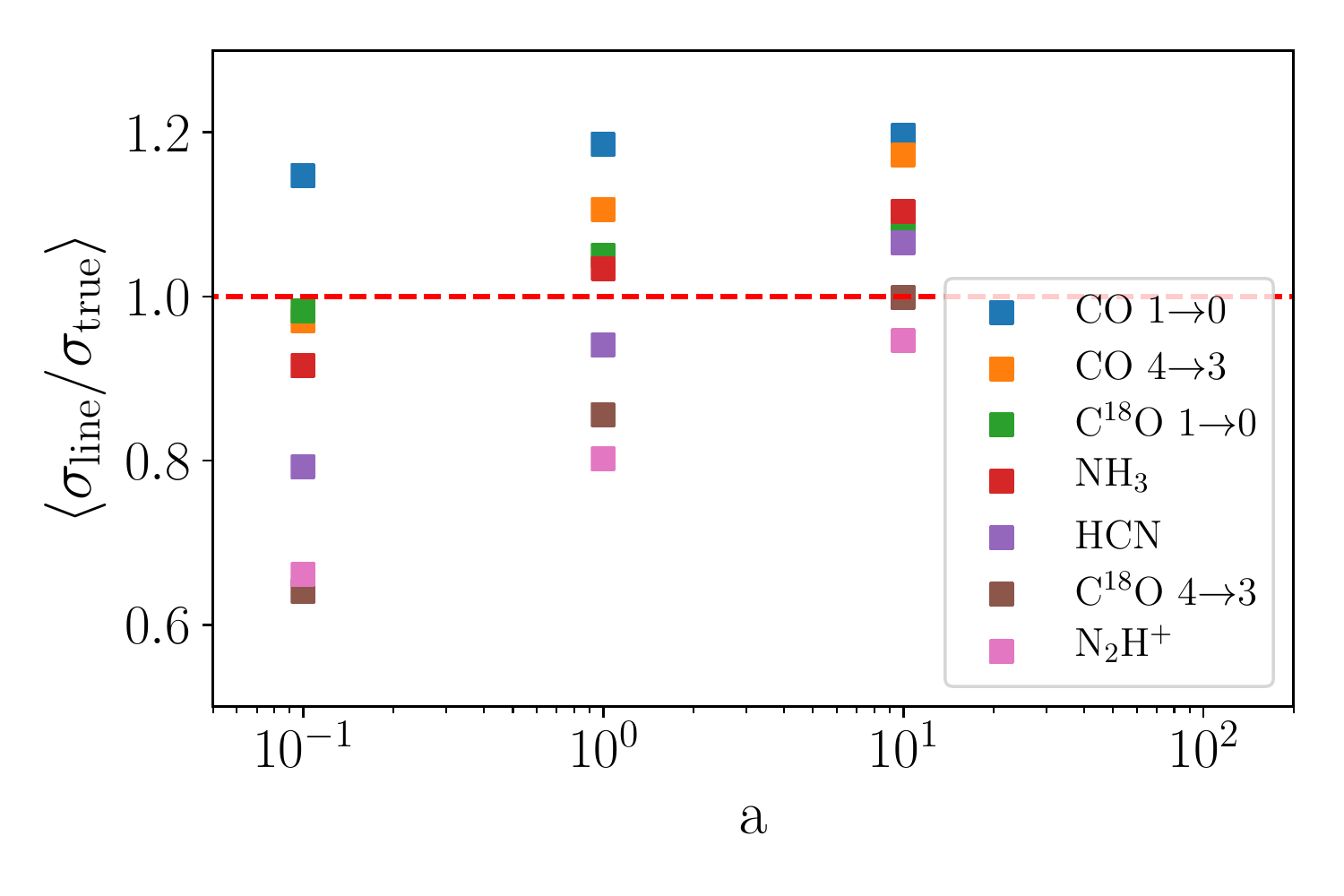}
\caption{Ratio of the luminosity-weighted mean velocity dispersion inferred from five emission lines as indicated in the legend, averaged over all simulations snapshots and orientations and normalised to the true velocity dispersion, as a function of the density scaling factor $a$.}
\label{fig:scaling}
\end{figure}

We consider $a=0.1$ and $a=10$ in addition to our standard case $a=1$, and generate PPV cubes and velocity dispersion measurements for all pixels in all snapshots following the same procedure described in \autoref{sec: methods} and \autoref{ssec: qualitative results}.  In \autoref{fig:scaling} we show the luminosity-weighted mean velocity dispersion inferred from all our molecular species, averaged over all simulations snapshots and orientations and normalised to the true velocity dispersion, versus the density scaling factor. We see that our conclusion that C$^{18}$O $1 \to 0$ is generally best, with NH$_3$ close behind, holds over a wide range of density, but that the amount of bias in these two species and in other tracers is density-dependent. Lower-density clouds suffer less opacity broadening and worse density bias, and thus make CO $1\to0$ closer to accurate and dense gas tracers further away from reality. Denser clouds have the opposite trend, suffering more opacity bias and less density bias, so that nearly any dense gas tracer works equally well, but CO is quite bad, with $\sim 30\%$ errors.

\subsection{Limitations of periodic boxes}

In addition to worrying whether our results depend on our choice of density scale, we can also worry that they depend on geometry. Our simulations are periodic boxes representing the central regions of molecular clouds, while real molecular clouds have dense material concentrated towards the center, surrounded by more diffuse molecular material toward the cloud's edge. It is therefore important to consider the extent to which our use of the periodic box approximation might affect our conclusions. \citet{Kowal07} have studied this question by comparing uniform-density periodic boxes such as ours to simulations in which an overall density gradient is applied on top of the periodic box, creating an effective boundary to the cloud. They find that the boundary of the molecular cloud increases the proportion of low density gas due to the disturbance of the diffuse ambient medium. This has the effect of increasing the amount of emission per unit total cloud mass from low-density tracers such as CO J=$1\to 0$, but does not affect the high density part of the density PDF, and thus has a small effect on high density tracers, particularly C$^{18}$O J=$4 \to 3$, HCN and N$_2$H$^+$. Thus the only line for which our results are potentially affected by our use of a periodic box is CO J=$1\to 0$. Moreover, the direction of the bias from observation of any particular tracer depends on the extent to which that tracer departs from the ``true'' mass distribution. The presence of a cloud boundary will change the ``true'' mass-weighted density PDF and the corresponding luminosity-weighted PDF of the tracers, but the correlation between the tracers and underlying mass is the same. Thus the direction of bias for CO is likely to be the same even in the presence of a cloud boundary. We therefore conclude that the main likely effect of adding a boundary layer to our cloud would be to change the absolute amount, but not the direction, of the bias for CO J=$1\to 0$. Other results would change minimally.

\section{Conclusions}
\label{con}

In this paper, we investigate the factors that drive differences in the linewidths of molecular clouds measured with various tracers. We carry out this investigation using a suite of self-gravitating MHD simulations of molecular clouds, covering a wide range of magnetic field strength and evolutionary state. For each of our sample simulations, we model the line emission using a large-velocity gradient approximation applied cell-by-cell to create synthetic PPV cubes that we use to investigate cloud kinematic structure in a variety of tracers. We specifically explore the effect of density-dependent emission and opacity broadening on observed linewidths, two mechanisms that have been discussed in the literature before, but never systematically investigated together. We also explore the effects of finite resolution and signal to noise ratio. The major findings of this paper are summarized below:

\begin{enumerate}
    \item Molecular lines that are sensitive to denser gas tend to produce systematically lower estimates of the gas velocity dispersion. This is a direct consequence of the linewidth-size relation obeyed by turbulent molecular clouds: tracers that are excited primarily in high-density gas tend to produce most of their emission from compact regions that, as a result of the linewidth-size relation, have small velocity dispersions and thus underestimate the true velocity dispersions of large clouds. Low-density tracers, by contrast, sample larger regions and therefore return larger velocity dispersions that are closer to the true velocity dispersion. 
    
    \item Opacity broadening also introduces a significant bias in the linewidths measured with optical thick tracers like CO $J=1\rightarrow 0$. The effect here tends to be opposite to the density bias: tracers that are easily excited in low-density gas, such as CO, tend to have high optical depths near line centre. This preferentially suppresses emission from the line centre, biasing inferred velocity dispersions too high. The relationship between optical depth and density-dependent excitation is complex, because high optical depth lowers effective critical density, while sub-thermal excitation can, depending on the molecule and line, either increase or decrease the optical depth. For CO, opacity broadening appears to be the more important effect, but which factors are dominant must be determined on a line-by-line basis.
    
    \item Bias induced by noise, finite spectral resolution and beam smearing from the telescope is mainly set by a competition between beam and noise effects. Noise introduces a bias whose effect is opposite that of opacity broadening, as it contaminates the line wings significantly, which artificially reduces the inferred linewidth; low density tracers are the most seriously affected. Beam-smearing, on the other hand, increases the linewidth slightly for high density tracers. At low SNR, the combined effects lower the linewidth of all tracers, while at high SNR, the linewidths of low density tracers are slightly reduced, and those of high density tracers are increased by a few percent due to beam-smearing.
    
    \item The competing biases of opacity broadening and density-dependent excitation lead to a ``sweet spot'' where, at fixed SNR, the overall bias is minimal, for three common tracers: the $J=4\rightarrow3$ transition of CO, the $J=1\rightarrow 0$ transition of C$^{18}$O and the $(1,1)$ inversion transition of NH$_3$. These lines generally produce the best estimates of true velocity dispersion for a typical molecular cloud, with errors below $\sim 10\%$ ($\sim 20\%$ for CO 4$\to$3). This statement is robust against variations of magnetic field strength, evolutionary state, and orientation relative to the direction of the overall magnetic field. By contrast, CO $J=1\rightarrow 0$ lines tend to produce velocity dispersions that are too large by $\approx 20\%$, while denser gas tracers such as HCN and N$_2$H$^+$ tend to underestimate the true velocity dispersion by similar amounts. However, these biases must be weighed against those produced by finite SNR, since the C$^{18}$O J=$1\to0$ and NH$_3(1,1)$ lines tend to be faint, and thus require longer integration times than for some other lines to reach SNR values high enough that noise does not dominate the uncertainty.
    
    \item The level of bias in various tracers is sensitive to the mean density of the region being observed. Over a wide range of density C$^{18}$O remains the best estimator of the true velocity dispersion, with NH$_3$ close behind,  but that the amount of bias in these two and in other tracers is density-dependent. In extreme cases, errors in the estimated velocity dispersion can be as large as 50\% high or low, depending on the cloud density and the choice of tracer.

\end{enumerate}

\section*{Acknowledgements}

MRK acknowledges funding from the Australian Research
Council through the Future Fellowship (FT180100375) and
Discovery Projects (DP190101258) funding schemes. BB acknowledges support Simons Foundation Flatiron Institute and the Center for Computational Astrophysics (CCA). Simulations used for this work are part of the Catalog for Astrophysical Turbulence Simulations (CATS) project hosted by CCA at \url{www.mhdturbulence.com}.   This work made use of resources from the National Computational Infrastructure (NCI), which is
supported by the Australian Government, through grant jh2.

%%%%%%%%%%%%%%%%%%%%%%%%%%%%%%%%%%%%%%%%%%%%%%%%%%



\bibliographystyle{mnras}
\bibliography{refs}

%%%%%%%%%%%%%%%%%%%%%%%%%%%%%%%%%%%%%%%%%%%%%%%%%%

%%%%%%%%%%%%%%%%% APPENDICES %%%%%%%%%%%%%%%%%%%%%

\appendix

\section{Dependence of results on different parameters}
\label{appd:full_par}
While we investigate the effect of various biases have on the linewidth for two example snapshots and orientation perpendicular to the magnetic field in the main body of the paper, in this appendix we explore the dependence of our conclusions on the following simulation parameters: plasma $\beta$, time (i.e., evolutionary state), and orientation (i.e., whether the line of sight is perpendicular or parallel to the mean magnetic field).

\begin{figure*}
\centering
  \includegraphics[width=1.\linewidth]{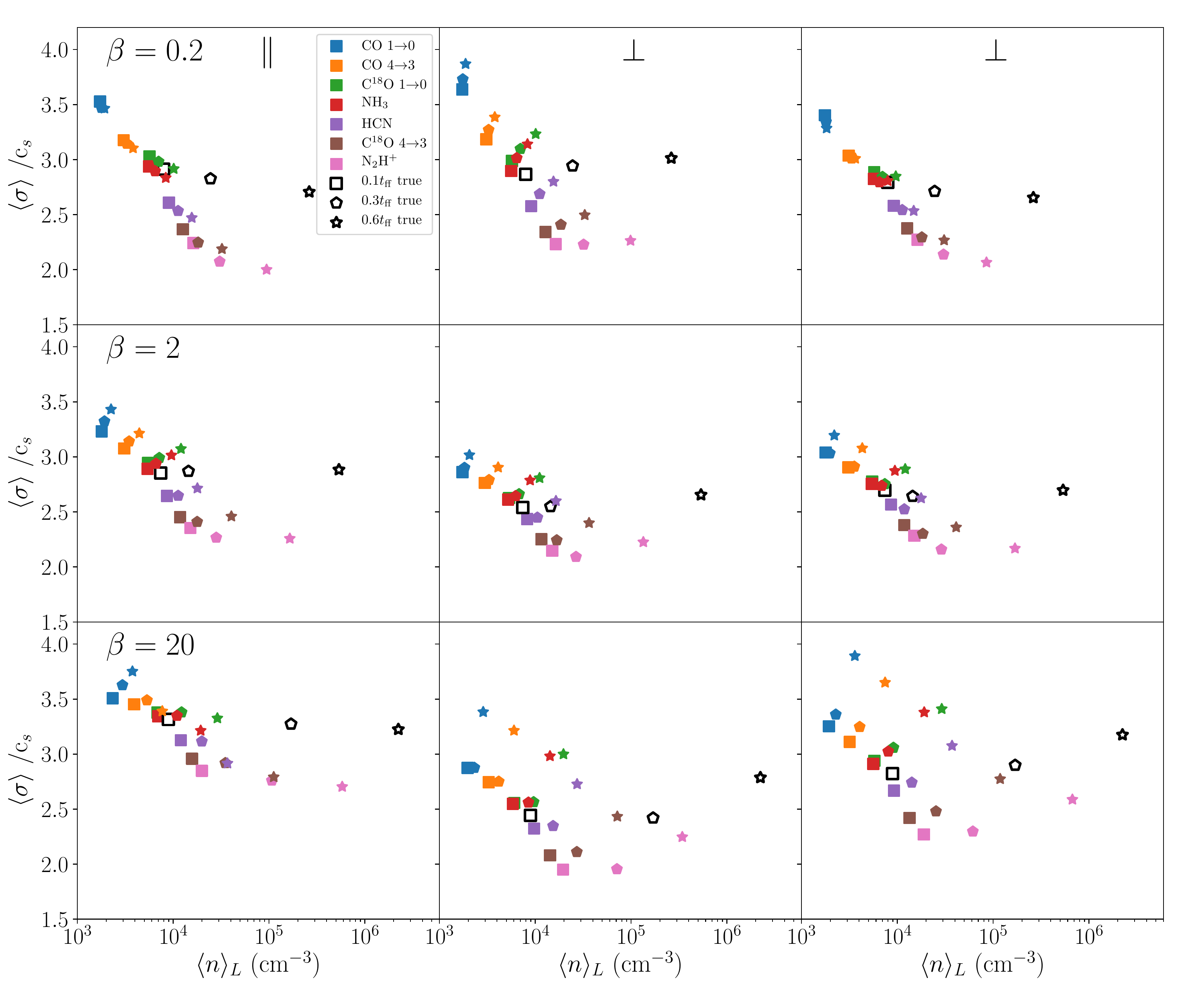}
\caption{Luminosity-weighted second moment vs.~luminosity-weighted mean density for all snapshots in all simulations. We show the second moment measured along the two cardinal axes perpendicular to the mean magnetic field in the centre and right columns, and along the axis parallel to the mean magnetic field in left column. From top to bottom, we show the $\beta = 0.2, 2$, and 20 runs. Points are colour-coded by time. Open symbols, labelled ``true'' in the legend, show the true mass-weighted mean density and velocity dispersion for each snapshot.
}
\label{fig:appd_2nd_m_vs_Lrho}
\end{figure*}

\begin{figure}
\centering
  \includegraphics[width=.8\linewidth]{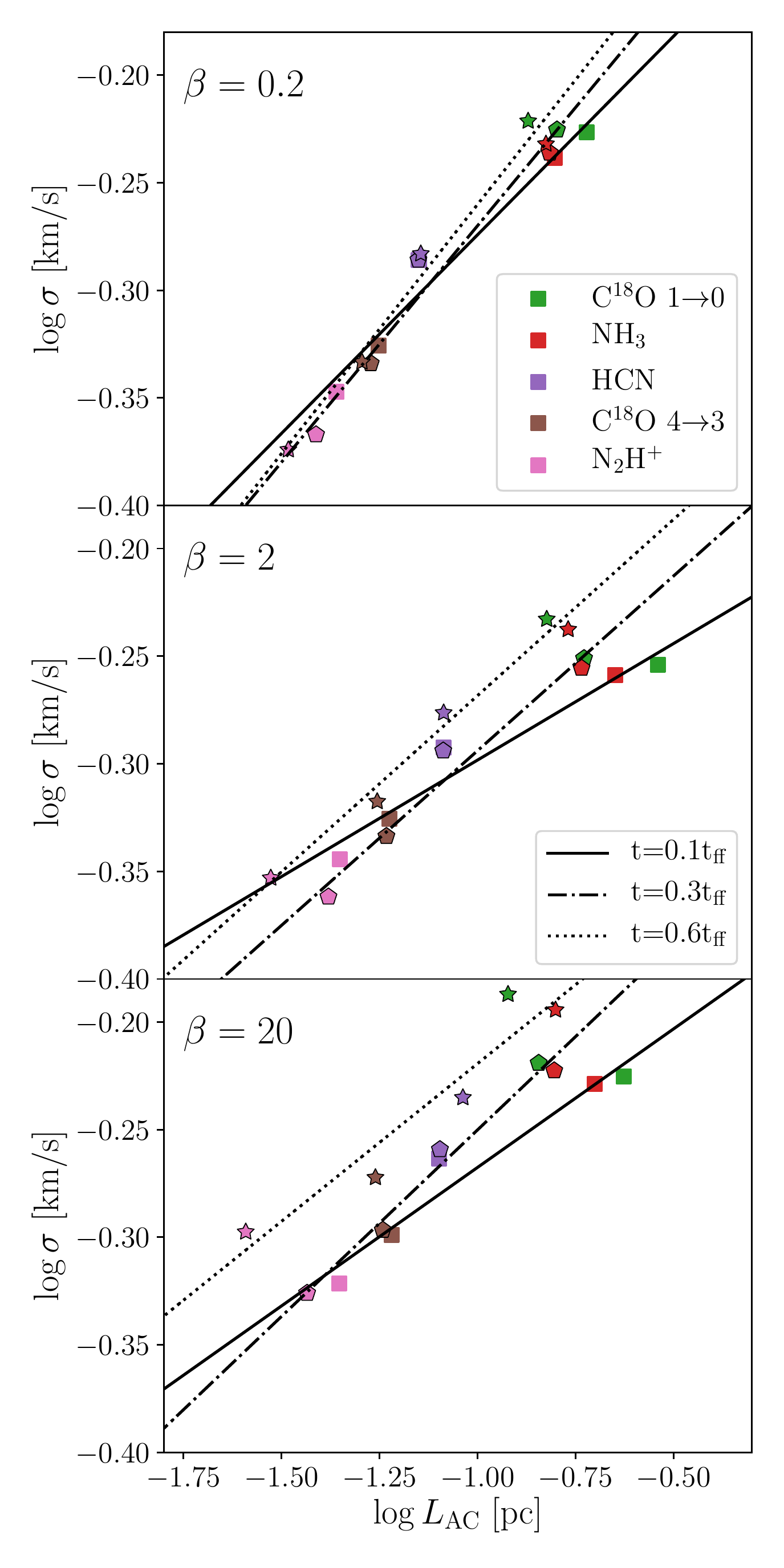}
\caption{Luminosity weighted second moment averaged over cardinal directions, $\sigma$, as a function of auto-correlation size of emitting regions $L_{\rm AC}$ for all snapshots. Lines show least squares linear fits to the data points at the time indicated in the legend. From top to bottom, we show the $\beta_0$ = 0.2, 2, and 20 runs. Note that this plot does not include CO, because we are unable to define $L_{\rm AC}$ for it. 
}
\label{fig:appd_LS_relation}
\end{figure}

\begin{figure*}
  \centering
  \includegraphics[width=\linewidth]{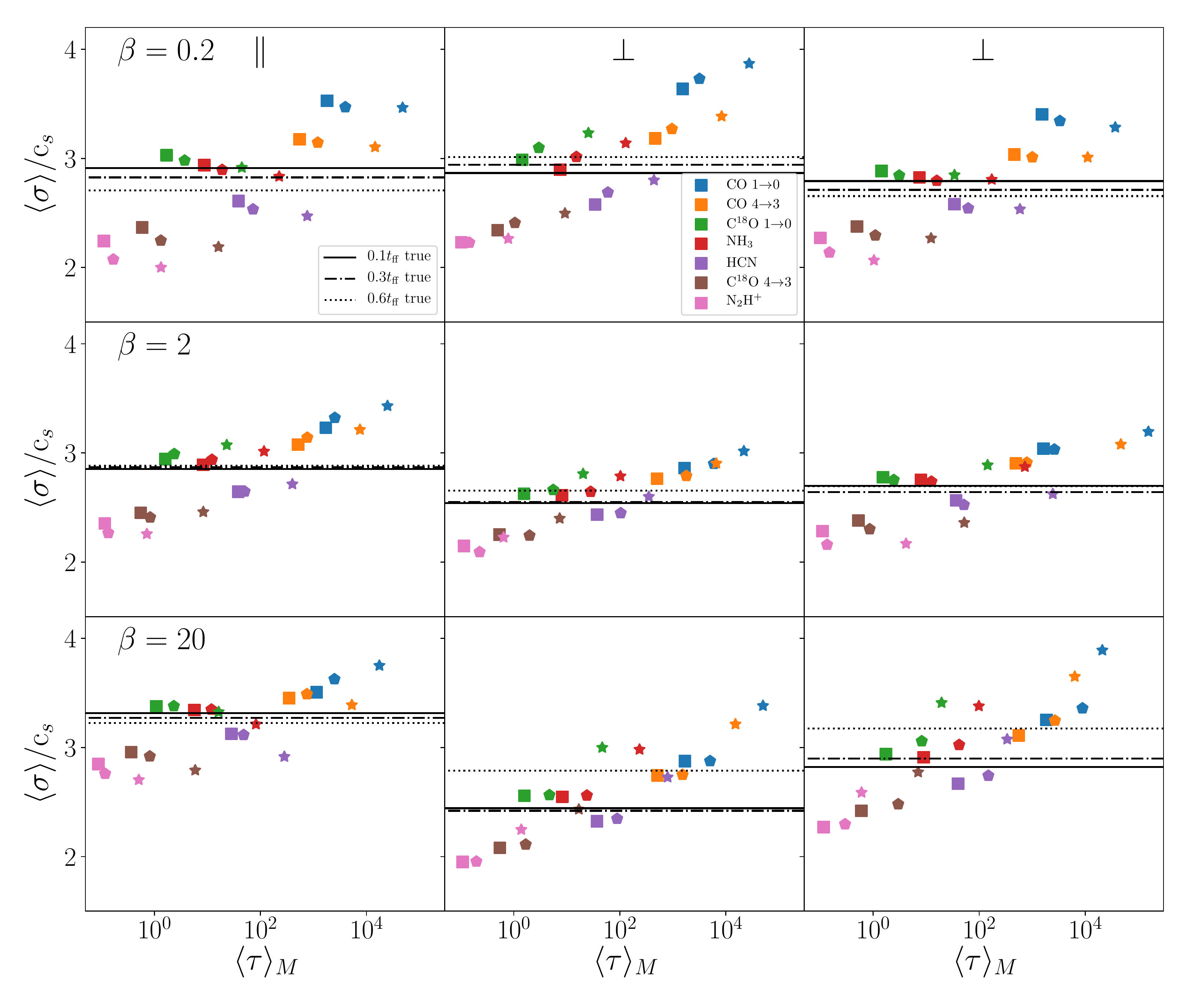}
\caption{Luminosity-weighted mean second moment versus mass-weighted mean opacity for all snapshots in all simulations. Dotted horizontal lines show the value of the true velocity dispersion at each time.}
\label{fig:appd_second_moment-opacity}
\end{figure*}

We first investigate whether the correlation between density and linewidth is sensitive to these parameters. In \autoref{fig:appd_2nd_m_vs_Lrho} we compare the linewidth with the luminosity-weighted mean density for all the snapshots and orientations. We see that the correlation remains essentially unchanged for all combinations of parameters. The greatest sensitivity is to evolutionary state, but even this dependence is weak. To go a step further, we compare the measured linewidth averaged over cardinal axis in each tracer with the corresponding characteristic emitting size in \autoref{fig:appd_LS_relation} for all snapshots. We see that the near-linear correlation between $\log \sigma $ and $\log L_{\rm AC}$ holds for all snapshots. We then show the dependence of linewidth on opacity for all snapshots and orientations in \autoref{fig:appd_second_moment-opacity}. Again we see that the general trend is similar to that shown in \autoref{fig:second_moment-opacity} for all combinations of parameters.

\begin{figure*}
  \centering
  \includegraphics[width=\linewidth]{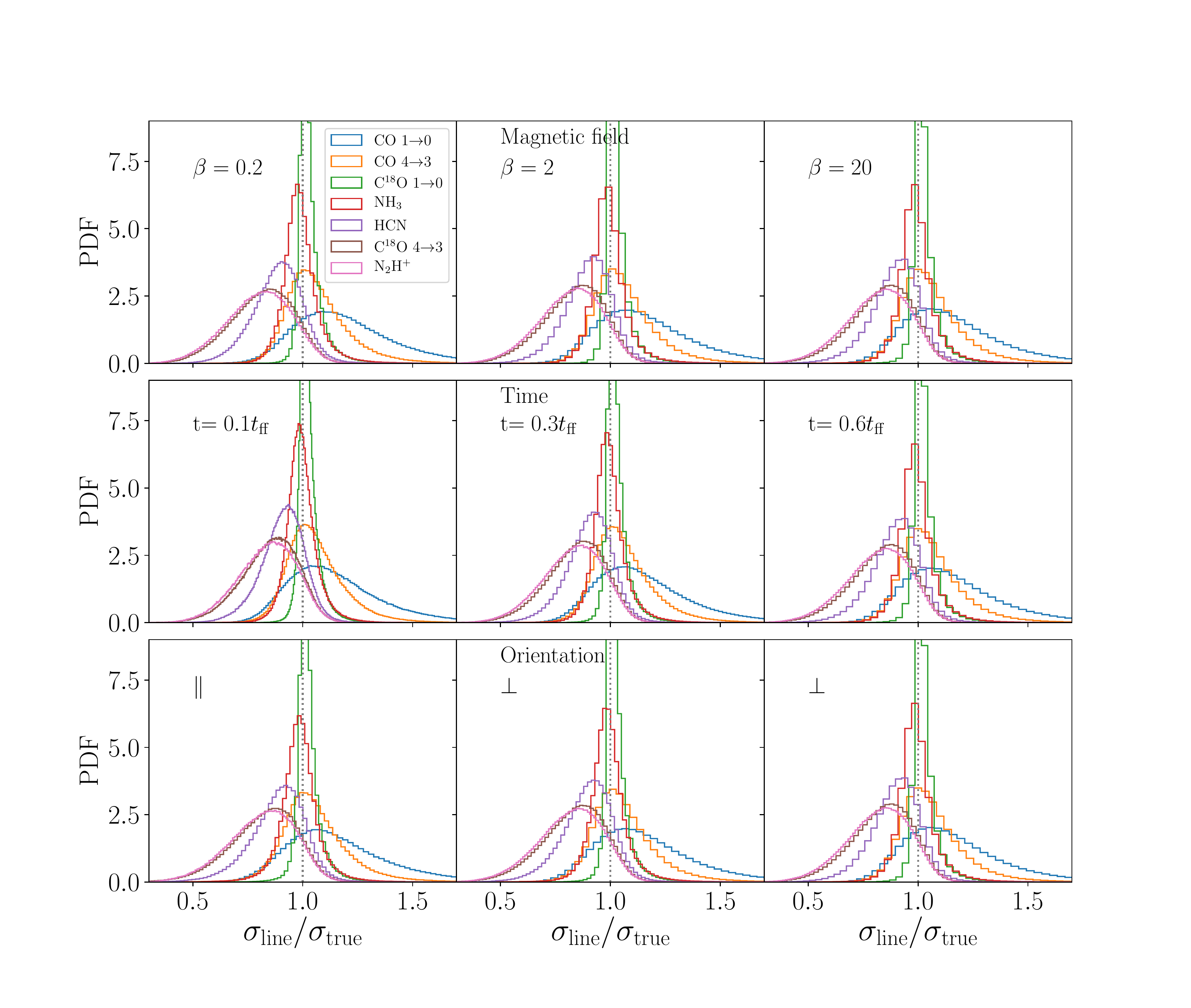}
\caption{Same as \autoref{fig:sigma_line_vs_sigma_true_all}, but rather than combining all snapshots of all simulations, the different rows show simulations binned by plasma $\beta$ (top row), evolutionary time (middle row), and orientation relative to the magnetic flux (bottom row). In each panel, the histogram shown is for all pixels in all simulation snapshots meeting the indicated condition, e.g., the panel labelled $\beta=0.2$ is the histogram of all pixels in all simulation snapshots and orientations for which $\beta=0.2$.
}
\label{fig:appd_sigma_line_vs_sigma_true}
\end{figure*}

Having illustrated that the main results in \autoref{sec: results} does not change qualitatively against different parameters. Our final step is therefore to determine whether our conclusions about which lines work best depends on the simulation parameters. \autoref{fig:appd_sigma_line_vs_sigma_true}
is the same as \autoref{fig:sigma_line_vs_sigma_true_all}
in that it shows distributions of $\sigma_{\rm line}/\sigma_{\rm true}$, but now with snapshots separated in bins of plasma $\beta$ (top row), simulation time (middle row), and orientation (bottom row). Surprisingly, we see that there are not any obvious variations in the distributions of $\sigma_{\rm line}/\sigma_{\rm true}$ with these parameters: in every case, C$^{18}$O and NH$_3$ are best, with errors below $\sim 10\%$,  CO is biased high by $\sim 20\%$, while the denser tracers are biased low by a similar amount.

%%%%%%%%%%%%%%%%%%%%%%%%%%%%%%%%%%%%%%%%%%%%%%%%%%

% Don't change these lines
\bsp	% typesetting comment
\label{lastpage}
\end{document}